\documentclass[11pt]{article}

\usepackage{amsmath}
\usepackage{amsfonts}
\usepackage{amssymb}
\usepackage{amsthm}
\usepackage{latexsym}
\usepackage{listings}

\numberwithin{equation}{section}
\newtheorem{axm}{A}

\newtheorem{crule}{CR}
\newtheorem{definition}{Definition}

\newcommand{\be}{\begin{equation*}}
\newcommand{\ee}{\end{equation*}}
\newcommand{\bal}{\begin{aligned}}
\newcommand{\eal}{\end{aligned}}
\newcommand{\ba}{\begin{eqnarray}}
\newcommand{\ea}{\end{eqnarray}}

\newcommand{\mb}{\mathbb}

\begin{document}

\title{Program Verification of Numerical Computation.}

\author{Garry Pantelis}

\maketitle

\abstract

These notes outline a formal method for program verification of numerical computation.
It forms the basis of the software package \emph{VPC} in its initial phase of development.
Much of the style of presentation is in the form of notes that outline the definitions and rules upon which \emph{VPC} is based.
The initial motivation of this project was to address some practical issues of computation, especially of numerically intensive programs that are commonplace in computer models. 
The project evolved into a wider area for program construction as proofs leading to a model of inference in a more general sense.
Some basic results of machine arithmetic are derived as a demonstration of \emph{VPC}.

\section{Introduction}

Mathematical and computer modelling is an important component of many scientific studies.
Applications in areas such as hydrodynamics employ computer models that are characterized by numerically intensive computation and are prime examples where program verification is desirable.
The main problem arises from the fact that machine numbers form a finite set so that any operation of numerical computation must ensure that the machine recognizes the output as a number.
Numerical computation failures manifest themselves in the form of underflows and overflows.

The objective here is to investigate a method for verification of programs that are largely based on numerical computation.
The objective is not to develop a tool that will take, as input, an arbitrary program and test the validity of that program.
Instead we can think of extracting from an arbitrary program the sequential order of arithmetic operations and test for computability.

In any attempt to construct a tool for program verification for numerical computation one first looks to the basic foundations of arithmetic starting with the axioms of rings and fields (see for example \cite{burk}, \cite{mclane}).
Unfortunately, when encountering machine arithmetic one will eventually observe a departure from the elementary rules of arithmetic upon which one has been accustomed to.
To explain some aspects of these departures one may delve deeper into analysis via topics such as modulo arithmetic and finite fields \cite{mclane}, but these too fall short of addressing many of the problems that are encoutered when dealing with machine arithmetic.

One promising approach is found in interval arithmetic \cite{moor}.
This is particularly useful for computations attempting to approximate continuum theories via floating point arithmetic but may play a less important role for discrete based models where exact rational or integer solutions are required.
To tackle this problem in its entirety one soon finds the need to investigate topics in a much wider area, many of which are found in the realm of computer science.
In particular, the initial motivation of program verification evolves into a wider area involving inference methods in a more general sense.  

Traditional studies of computers and computation often start by constructing a theoretical model that reflects some aspects of real world computers.
Such examples can be found in Turing machines along with theoretical abstractions of programming languages themselves such as lambda calculus \cite{baren} leading to the study of logic and the important link between programs and proofs.
The latter in turn leads one into the area of proof theory.
This is a wide area of study of which an excellent coverage can be found in \cite{buss}.

The approach taken here is to construct a formal language such that the rules of inference are dictated not so much by an external theory of computation and logic but rather by the constraints dictated by finite memory storage and allowable operations on a real world classical computer.
As a consequence there will be a need to abandon some of the expressiveness found in models based on current proof theory.  
The language is presented in a form that is less abstract than traditional studies of theoretical computers and is constructed in a way that is readily identifiable as a more practical guide to program verification through functional programming.
These methods will be described in the context of the software package \emph{VPC} (Verification of Program Computation) that is in its initial phase of development.

\vspace{5mm}
\noindent
\textbf{Sequential ordering.} 
Let $Q_i,~i=1,2,\ldots,n$, be statements. Consider the composite statement 
\be
[Q_1,\ldots,Q_n]
\ee
The list of statements is to be read in sequential order from left to right and the procedure halts after the reading of the last statement $Q_n$.
In a machine environment the $Q$'s may be instructions such as type checking operations or assignments.
The reading of the list $[Q_1,\ldots,Q_n]$ may halt prematurely if the machine encounters a statement that results in an execution error.   

The sequential ordering of statements in the list $[Q_1,\ldots,Q_n]$ is a key procedure of programming languages.
At first glance the sequential ordering is dictated by the following properties.

\begin{itemize} 

\item The order in which each statement appears in the list $[Q_1,\ldots,Q_n]$ is important although some interchange is possible under special conditions.

\item Each statement in the list $[Q_1,\ldots,Q_n]$ may have some kind of dependence on one or more statements that preceed it but no statement can have a dependence on a statement that follows it.

\end{itemize}
To understand how these two properties are linked we will need to associate with each statement a set of variables that act as either input or output parameters of that statement.
We may also consider instructions that essentially split a program into a number of parallel sequentially ordered streams.
This process is related to logical statements involving disjunctions.
This will be discussed further in a later section but for now it will suffice to consider programs defined by a single sequential stream as defined above.

Programs can be constructed by either imperative or functional programming languages.
Programs constructed by an imperative language will not be discussed in detail other than to acknowledge that they will form the collection of atoms of functional programs from which larger functional programs can be constructed.
Inference methods will be conducted in the setting of functional programs.
It will be seen that after introducing a set of rules, the process of constructing programs for program verification can be directly linked to inference methods.
While this will open up an opportunity to study a much wider area of applications, we will remain focussed on issues related to machine arithmetic.

\section{Types.}

We deal with objects and types.
Each object has a type.

\vspace{5mm}
\noindent
\textbf{Properties of types.}

\begin{itemize}

\item Types will be denoted by the symbols $\mb{A},\mb{B},\ldots,\mb{Z}$.
We may also attach to these symbols subscripts and primes.

\item Object $a$ has type $\mb{A}$ is denoted by $a:\mb{A}$.

\item An object may also be dependent on another object.
We write $a(n)$ to mean that the object $a$ depends on the parameter or object $n$.
Sometimes we may index a collection of objects using subscripts, e.g. $a_1,\ldots,a_n$ denotes a collection of objects that may have different types.

\item Types may be subtypes of types.
$\mb{A}$ is a subtype of $\mb{B}$ is denoted by $\mb{A}<:\mb{B}$.
Subtypes have the properties: (i) if $a:\mb{A}$ and $\mb{A}<:\mb{B}$ then $a:\mb{B}$,
(ii) if $\mb{A} <: \mb{B}$ and $\mb{B} <: \mb{C}$ then $\mb{A} <: \mb{C}$.

\item Types may also be dependent on objects.
We write $\mb{A}(a_1,\ldots,a_n)$ to mean that type $\mb{A}$ depends on the parameters or objects $a_1,\ldots,a_n$.
Parameter dependent types are subtypes of its generic type, i.e. $\mb{A}(a_1,\ldots,a_n)<:\mb{A}$

\end{itemize}

\section{Alphabet and strings.}

Here we shall work in a machine environment based on a classical computer.
An important feature of a real world machine environment is the property of finite information storage along with a collection of well defined operations.
The context is defined by the following machine specific parameters.
\be
\begin{array}{ll}
K & \text{number of characters in the alphabet.} \\
L & \text{maximum number of characters in any string.} \\
M & \text{maximum number of elements of any list stored as an array.} \\
\end{array}
\ee
We start by defining the alphabet $\mb{J}$ as a collection of symbols or characters
\be
s(1),\ldots,s(K)
\ee
The alphabet $\mb{J}$ will consist of the following characters.

\begin{itemize}

\item Letters (upper and lower case).
\be
\bal
a~b \ldots z \\
A~B \ldots Z \\
\eal
\ee

\item Digits.
\be
1~2~3~4~5~6~7~8~9~0
\ee

\item Special characters.
\be
~.~,~+~-~*~/~(~)~[~]
\ee

\end{itemize}

\begin{definition}(String.)

\begin{itemize}

\item A string of the alphabet $\mb{J}$ is a sequence of characters $s(i_1) s(i_2) \ldots s(i_j)$, $1 \leq i_1,\ldots,i_j \leq K$, $1 \leq j \leq L$.

\item A string is given a type denoted by $\mb{S}$.

\item There are two main subtypes of strings.

\be
\begin{array}{ll}
\mb{C}<:\mb{S} & \text{alphanumeric strings comprised of any combination of letters and digits} \\
& \text{with the first character always being a letter.} \\
 \mb{I}<:\mb{S} & \text{signed integers comprised of digits preceeded by a sign $\pm$.} \\
\end{array}
\ee

\end{itemize}

\end{definition}

\vspace{5mm}
\noindent
\textbf{Alphanumeric strings.}
Alphanumeric strings are assigned the type $\mb{C}$ and are often used to represent names of programs and variable names of elements of input/output lists of programs.
Variable names of elements of input/output lists of programs serve as place holders for the assigned values that are defined as specific types within the program.
We write $a:\mb{C}$ to stress that $a$ is a dummy variable that represents an alphanumeric string.
Upon entry to a program we may also write $a:\mb{A}$ to denote that the alphanumeric string has been assigned a value of type $\mb{A}$.
The assigned value may be an integer or another alphanumeric string.
 
If $a$ and $b$ are dummy variables representing two alphanumeric strings we write $a=b$ to denote that the the two alphanumeric strings are identical.
The sense in which the equality is being used here will often appear in the context of definitions related to program structures and properties.

We may also write $a=b$ to denote that the assigned value of the alphanumeric variable represented by the dummy variable $a$ is identical to the assigned value of the alphanumeric variable represented by the dummy variable $b$.
These assigned values may be numbers or other alphanumeric strings of a specific subtype and the equality is always accompanied with type checking statements of the assigned values.
The sense in which the equality is used here will always be stated to avoid confusion.
For two alphanumeric variable names represented by $a$ and $b$ we write $a:=f(b)$ to denote that the value assigned to $a$ is acquired through the assignment function $f$ acting on the value assigned to $b$.

\vspace{5mm}
\noindent
\textbf{Machine numbers.}
An object of type $\mb{I}$ is a string that can be assigned any one of the integer values
\be
0, \pm 1,\ldots, \pm N ,
\ee
where $N$ is the maximum positive integer and is a specific machine parameter.
We shall make extensive use of the following subtypes of $\mb{I}$.
\be
\begin{array}{ll}
\mb{I}_0 & \text{$a:\mb{I}_0$ denotes $a:\mb{I}$ and $0 \leq a \leq N$} \\
\mb{I}_+ & \text{$a:\mb{I}_+$ denotes $a:\mb{I}$ and $0 < a \leq N$} \\  
\end{array}
\ee
We adopt the usual convention of dropping the prefex $+$ sign when dealing with positive integers.

Our main objective here is to describe the software package \emph{VPC} as a tool for verification of numerical computation.
For the purpose of demonstration only we will restrict the outline to machine integer arithmetic but it should be kept in mind that \emph{VPC} has a much wider area of application that includes floating point arithmetic. 

\section{Lists.}

\textbf{Type.}
\be
\begin{array}{ll}
\mb{L} (n) & \text{type list with $n:\mb{I}_0$ elements.} \\
\mb{L} & \text{type generic list of unspecified length, $\mb{L} (n) <: \mb{L}$.} \\
\end{array}
\ee

\vspace{5mm}
\noindent
\textbf{Properties of lists.}

\begin{itemize}

\item Lists contain elements of strings. A list $x : \mb{L}(n)$, has the representation
$x=[x_1,\ldots,x_n]=[x_i]_{i=1}^n$, where $x_i : \mb{S}$, $n:\mb{I}_0$.
The object $n$ is referred to as the length of the list $x$.
The notation $x_i \in x$ means that $x_i$ is a element of the list $x$.

\item An empty list $x:\mb{L}(0)$ is denoted by $x=[~]$.
If, under the list representation $x=[x_i]_{i=1}^n$, we have $n=0$ then it is understood that $x$ is the empty list. 

\item $x=[x_1,\ldots,x_n]$ is a list with elements $x_1,\ldots,x_n$ while $[x]$ is a list with a single element $x$.

\item In any list $x=[x_1,\ldots,x_n]$, all elements are strings, i.e. of type $\mb{S}$, but elements of a list may be assignd values with different types.

\item In a list $x=[x_1,\ldots,x_n]$ any element of $x$ could itself be a list.
A list may sometimes be treated as a string using the hierachy of subtypes $\mb{L}<:\mb{S}$.
A list may also be treated as an array of strings.

\item \textbf{List equality.} If $a:\mb{L}(n)$ and $b:\mb{L}(n)$, $n:\mb{I}_0$, and $b_i=a_i$, $i=1,\ldots,n$, we write $a=b$.
We use equality for both senses of identity of alphanumeric variable names and the values assigned to the alphanumeric variable names.
Throughout, unless otherwise stated, equality will be assumed to be in the sense of the former, i.e. in the sense of identity of alphanumeric names.
Whenever the equality is used in the sense of assigned values it will be stated as such.

\end{itemize}

\vspace{5mm}
\noindent
\textbf{List operations.}

\begin{itemize}

\item \textbf{Empty list extraction.} Suppose that $a=[a_i]_{i=1}^n:\mb{L}(n)$, $n:\mb{I}_+$, contains a element $a_k \in a$ that is an empty list, i.e. $a_k=[~]$.
We may extract the empty list element and write  
\be
a=[a_1,\ldots,a_{k-1},a_{k+1},\ldots,a_n]
\ee
After empty list extraction we can automatically redefine $a:\mb{L}(n-1)$. 

\item \textbf{List concatenation.} If $a=[a_1,\ldots,a_m]$ and $b=[b_1,\ldots,b_n]$ are two lists then the concatenation of $a$ and $b$ yields the list $c:\mb{L}(m+n)$ given by
\be
c=[a,b]=\big [ [a_1,\ldots,a_m],[b_1,\ldots,b_n ] \big ] = [a_1,\ldots,a_m,b_1,\ldots,b_n ]
\ee 
The internal square brackets that deliminate the lists $a$ and $b$ may be removed.
The exception to this rule are I/O lists discussed below. 
If $x_1,\ldots,x_n$ are lists with representations $x_i=[x_{ij}]_{j=1}^{m_i}$, $i=1,\ldots,n$, then their list concatenation, z, is given by
\be
\bal
z = & [ \ldots [ x_1,x_2 ], x_3 ], \ldots ], x_n ] \\
= & [x_1,x_2,\ldots,x_n] \\
= & [x_{11}, \ldots ,x_{1m_1}, \ldots,x_{n1}, \ldots ,x_{1m_n}]
\eal
\ee 

\item \textbf{List intersection.} If $a=[a_1,\ldots,a_m]$ and $b=[b_1,\ldots,b_n]$ then the list intersection of $a$ and $b$, yields a new list $c:\mb{L}(k)$, $k \leq \min(m,n)$, where $c=[c_1,\ldots,c_k]$ contains all of the elements that are common to both $a$ and $b$.
We write $c=a \bigcap b$ to mean that $c$ is the list intersection of $a$ and $b$.
Whenever a list intersection is constructed the sequential order of the elements of $c$ are in the same hierachy of the sequential order as they appear in $a$. 
If $x_1,\ldots,x_n$ are lists then their list intersection, z, is given by
\be
\bal
z = & \bigcap_{i=1}^n x_i \\
= &  ( \ldots ( x_1 \bigcap x_2 ) \bigcap x_3 ) \bigcap \ldots ) \bigcap x_n ) \\
\eal
\ee 

\item \textbf{Removal of repeated elements of a list.} If $a=[a_1,\ldots,a_m]$ has repeated elements we can construct a new list $b=[a_{i_1},\ldots,a_{i_n}]$, $i_1 <~i_2 \ldots <~i_m$, by removing repeated elements as follows.
Reading the list $a$ from left to right, whenever a element is encountered that coincides with a preceeding element of $a$ then that element is extracted.
In other words, each element of $b$ contains all non repeated elements of $a$ and the first occurance of a repeated element of the list $a$, as read from left to right, maintining the order in which they appear in $a$. 
We write $b \simeq a$ to mean that $b$ is obtained by extracting repeated elements of $a$ by this procedure.

\item \textbf{List element substitution.} From the list $a=[a_1,\ldots,a_n]$ a new list
\be
b=[a_1,\ldots,a_{i-1},x,a_{i+1},\ldots,a_n]
\ee
is constructed by replacing the element $a_i$ in the list $a$ by $x$.
We write $b=a|_{a_i \to x}$.

\item \textbf{List element extraction.} Suppose that $a:\mb{L}(m)$ and $b:\mb{L}(n)$, $n \leq m$, such that all elements of $b$ are contained in the list $a$, i.e. $a \bigcap b = b$.
We can construct a new list $c$ obtained by the extraction from $a$ those elements found in $b$.
The new list maintains the sequential order found in $a$, i.e. $c=[a_{i_1},\ldots,a_{i_k}]$, $i_1 <~i_2 \ldots <~i_k$, $k \leq m-n$, where $a_{i_1},\ldots,a_{i_k}$ are all of the elements of $a$ not found in $b$.
We write $c=a \setminus b$ to denote the new list constructed in this way.

\end{itemize}

\vspace{5mm}
\noindent
\textbf{Sublists.}
Because of its importance, the notion of a sublist affords a more formal definition. 

\begin{definition}(Sublist.)
A list $b:\mb{L}(m)$, $m:\mb{I}_+$, is a sublist of $a=[a_1,\ldots,a_n]$, $m \leq n$, if it has the representation $b=[a_{i_1},\ldots,a_{i_m}]$, where $[i_1,\dots,i_m]$ is as an $m$-permutation of $[1,\ldots,n]$.
We write $b \subseteqq a$ to mean that $b$ is a sublist of $a$.
There are two cases that need to be distinguished.

\begin{itemize}

\item $b \subseteqq a$ and $m<n$. We say that $b$ is a strict sublist of $a$ and write $b \subset a$.

\item $b \subseteqq a$ and $m=n$ we either have $b=a$ or $b$ is a permutation of $a$. In either case we also have $a \subseteqq b$. 

\end{itemize}
The empty list $[~]$ is regarded to be a sublist of all lists.

\end{definition}

\vspace{5mm}
\noindent
\textbf{I/O Lists.}
An I/O list will refer to a list that appears as an input and output list of a program.
Whenever a list appears as an input or output list of a program it should not be read with the properties of the standard list concatenation as described above.
To illustrate this, consider the collection of lists $x_1,\ldots,x_n$ with representations $x_i=[x_{ij}]_{j=1}^{m_i}$, $i=1,\ldots,n$.
An I/O list $[x_1,x_2,\ldots,x_n]$ should be regarded as an object of type $\mb{L}(n)$ whose elements are distinct objects of type $\mb{L}(m_i)$, $i=1,\ldots,n$.
The list $[x_1,x_2,\ldots,x_n]$ of a standard list concatenation is an object of type $\mb{L}(m)$, where $m=\sum_{i=1}^n m_i$.
It is important to note that the list concatenation restriction only applies when a list appears as a program input/output list.
Otherwise the standard list operations on $x_1,\ldots,x_n$ apply. 

\section{Program structure.}

\vspace{5mm}
\noindent
\textbf{Program type and naming convention.}
Programs will be assigned a type denoted by $\mb{P}$.
Program names are assigned the type $\mb{P}_{name}$ and are a specific subtype of alphanumeric strings, i.e. $\mb{P}_{name} <: \mb{C}$.
The convention used here is to list all program names in the form of a string as a combination of letters and digits, with the starting character being an upper case letter and any following letters in lowercase. Programs are defined recursively as follows.

\begin{definition}(Program.)
A program has the representation $P(x,y)$ with the allocation of types of its component parts given by
\be
\begin{array}{ll}
P(x,y):\mb{P} & \text{program} \\
P:\mb{P}_{name} & \text{program name} \\
x:\mb{L} & \text{input list} \\
y:\mb{L} & \text{output list} \\
\end{array}
\ee

A program $P(x,y)$ satisfies all of the following conditions.

\begin{itemize}

\item Elements of the input and output lists are alphanumeric variable names (type $\mb{C}$) that serve as placeholders for assigned values.
The type of the assigned values of every element of the input and output lists of a program are checked within the program.

\item The variable names of the elements of the output list $y$ are distinct.

\item No element of the input list $x$ can have a variable name that coincides with a variable name of an element of the output list $y$, i.e. $x \bigcap y = [~]$.

\end{itemize}

\noindent
A program $P(x,y):\mb{P}$ can be represented by a list $P(x,y) = [P_i(x_i,y_i)]_{i=1}^n$, for some $n:\mb{I}_+$.
A program list satisfies all of the following conditions.

\begin{itemize}

\item $P_i(x_i,y_i):\mb{P}$, $i=1,\ldots,n$.

\item $\bigcap_{i=1}^n y_i = [~]$.

\item (I/O dependency condition) $x_i~\bigcap~[y_j]_{j=i}^n = [~]$, $i=1,\ldots,n$, where $[y_j]_{j=i}^n$ is a standard list concatenation.

\item $P \neq P_i$, $i=1,\ldots,n$.

\item The output list $y$ is defined by the standard list concatenation
\be
y = [ y_i ]_{i=1}^n
\ee

\item The input list $x$ is defined by
\be
x = x^\prime \setminus (x^\prime \bigcap y) , \quad x^\prime \simeq [x_i]_{i=1}^n
\ee
where $[x_i]_{i=1}^n$ is the standard list concatenation of the lists $x_1,\ldots,x_n$.

\item The execution of the program $P(x,y)$ is completed when the execution of all element programs of the list are completed in the sequential order from left to right.  

\end{itemize}
In the representation of a program list $P(x,y) = [P_i(x_i,y_i)]_{i=1}^n$, $P(x,y)$ is called the main program and each $P_i(x_i,y_i)$ is called a subprogram of $P(x,y)$.
The empty program is denoted by $[~]$.

\end{definition}

\vspace{5mm}
\noindent
\textbf{Computability.} 
In all programs the type of the assigned values of all the elements of the input and output lists are checked within the program.
Type violations are associated with execution errors.    

\begin{definition}(Execution error.)
A program $P(x,y):\mb{P}$ will halt prematurely with an execution error if during the execution of $P(x,y)$ there is a type violation of any assigned value of the elements of its input or output list.

\end{definition}

\begin{definition}(Computability.)
A program $P(x,y):\mb{P}$ is said to be effectively computable, or simply computable, with respect to the values assigned to the input list $x$, if upon execution it does not halt with an execution error and returns the output $y$.
\end{definition}

\vspace{5mm}
\noindent
\textbf{Programs as vertical lists.}
A program with the list representation $P(x,y)=[P_i(x_i,y_i)]_{i=1}^n$ can be represented as a vertical list

\be
P(x,y) = \left \{ 
\begin{array}{l}
P_1(x_1,y_1) \\
\hspace{5mm} \ldots \\
P_n (x_n,y_n) \\
\end{array}
\right .
\ee
We shall regard the above vertical and horizontal lists of a program $P(x,y)$ to be just different representations of the same program.
When $n$ is relatively small it is sometimes useful to write programs as horizontal lists and regard them as having type $\mb{S}$.
When $n$ is relatively large it is more convenient to store programs as arrays and display them as vertical lists.

\vspace{5mm}
\noindent
\textbf{Notes.}

\begin{itemize}

\item A program $P(x,y)$ can be regarded as a list $P(x,y) = [P_i(x_i,y_i)]_{i=1}^n$, $n:\mb{I}_0$.
For $n=1$ we simply drop the list representation and write $P(x,y)$.
The execution of the program $P(x,y)$ is completed when all of the subprograms $P_i(x_i,y_i)$, $i=1,\ldots,n$, have been executed in the sequential order from left to right in the program list.

\item A program list $P(x,y) = [P_i(x_i,y_i)]_{i=1}^n$ should be thought of as being a core program imbedded in a larger program.
In this larger program the core program is preceeded by a subprogram that reads in data and assigns values to alphanumeric variable names as well as assigning the type of the assigned values.
Following the core program is a subprogram that writes the assigned values of the output to a file and/or screen.
For the purpose of analysis the core program $P(x,y)$ may be considered in isolation.

\item By definition the variable names of the elements of the output list of all subprograms $P_i(x_i,y_i)$ of a program list $P(x,y) = [P_i(x_i,y_i)]_{i=1}^n$ must be distinct.

\item By definition the input list $x_i$ of each subprogram $P_i(x_i,y_i)$ of a program list $P(x,y) = \big [P_i(x_i,y_i)]_{i=1}^n$ must not depend on any element of the output list $y_i$ or any element of an output list of subprograms that follow $P_i(x_i,y_i)$ in the program list.
This disallows reassigning values to a variable name as is a common practice in imperative programming.

\item If the empty program is encountered in the execution of a program list then the program does not halt and execution proceeds to the next subprogram of the list.

\item Subprograms of a program list that are empty programs can be immediately removed by the process of an empty list extraction.

\item Given a program $P(x,y) = [P_i(x_i,y_i)]_{i=1}^n$, the output $y$ of a main program is a standard concatenation of the output lists of its subprograms.
The input list of any subprogram can contain elements that are also elements of output lists of the subprograms that preceed it in the program list.
Some of these internal outputs are sometimes not of interest to the application for which the main program is designed and are often regarded as free parameters.
It is common practice in imperative programming to regard free parameters as having utility for internal purposes only and discarded upon the execution of the main program .  
For functional programming we shall not make use of the notion of free parameters and include in the output list of the main program all of the elements of the output lists of its subprograms.
This may lead to an accumulation of a large number of variables that need to be stored in memory but there are advantagses to this approach.
Firstly, the removal of the notion of free parameters will avoid the need to introduce some cumbersome details in the definitions of the rules of program construction.   
Secondly, if the main program, $P(x,y)$, is later embedded as a subprogram into another program list the elements of $y$ that may have otherwise been discarded as free parameters can sometimes be used as input by a subprogram in the new list.

\end{itemize}

\section{Atomic programs.}

Functional programs are built up from lists of atomic programs.
Atomic programs are to be understood as being constructed from some imperative language.
The imperative program list of atomic programs will not be presented, only their functionality will be defined.
The functionality of atomic programs that are used in this paper are defined in the Appendix.

\begin{definition}(Atomic program.)
Atomic programs are a subtype of program type, $\mb{P}_{atomic}<:\mb{P}$.
An atomic program $P(x,y):\mb{P}_{atomic}$ must include type checking for the assigned values of every element of the input and output lists.
If for any element of the input and output lists there is a type violation the program halts prematurely as a type violation error.
Otherwise the atomic program returns the output $y$, where $y$ may be the empty list.
Atomic programs may call other atomic programs but each atomic program introduces a new functionality.
\end{definition}

\noindent
Atomic programs can be grouped into the three subtypes of type checking, type assignment and value assignment.

\begin{definition}(Type checking programs.)
A type checking program $P(x,y):\mb{P}_{type}$ is an atomic program, $\mb{P}_{type} <: \mb{P}_{atom}$, with the following properties.

\begin{itemize}

\item The output list $y$ is the empty list so that type checking programs have the representation $P(x,[~])$.

\item The type of the assigned values of every element of the input list is checked upon entry.

\item If there is a type violation the program halts prematurely with a type violation error.

\end{itemize}

\end{definition}

\begin{definition}(Type assignment programs.) 
A type assignment program $P(x,y):\mb{P}_{tassign}$ is an atomic program, $\mb{P}_{tassign} <: \mb{P}_{atom}$, with the following properties.

\begin{itemize}

\item The output list $y$ is the empty list so that type assignment programs have the representation $P(x,[~])$.

\item The type of the assigned values of every element of the input list is checked upon entry.
The check is performed on the type already assigned to the variable upon entry and not the type that is to be assigned.

\item If there is a type violation the program halts prematurely with a type violation error.

\item If, upon entry, there are no type violations a type assignment program then assigns the new type to elements of the input list that are the target of that type assignment program.

\item Once a variable is assigned a new type it is internally stored in memory so that if the variable is encountered as input of a following subprogram of a program list it retains that assigned type.
 
\end{itemize}
If $v$ is a dummy variable representing an alphanumeric name of an element of the input list $x$ such that the value assigned to $v$ has type $v:\mb{V}$ upon entry then we write $v::\mb{U}$ to denote that the assigned value of $v$ has been assigned the type $\mb{U}$.

\end{definition}

\begin{definition}(Value assignment programs.)
A value assignment program $P(x,y):\mb{P}_{assign}$ is an atomic program, $\mb{P}_{assign} <: \mb{P}_{atom}$, with the following properties.

\begin{itemize}

\item Programs of type $\mb{P}_{assign}$ have a dual function of type checking and value assignment.

\item The type of the assigned values of every element of the input list is checked upon entry.

\item If there is a type violation the program halts prematurely with a type violation error.

\item If there are no type violations a value assignment program attempts to assign a value to each element of the output list through the action of an assignment function.

\item We write $y:=f(x)$, where $f$ is the assignment function of $P(x,y)$.
If $y$ has the list representation $y=[y_1,\ldots,y_n]$ then the assignment function $f(x)$ has the list representation $f(x)=[f_1(x),\ldots,f_n(x)]$.

\item The type of the value assigned to each element of the output list through the acion of the assigment function is checked.

\item If there is a type violation of an assigned value of an element of the output list the program halts prematurely with a type violation error.
If there are no type violations the value assigned list $y$ is returned as output.

\end{itemize}

\end{definition}

\vspace{5mm}
\noindent
\textbf{Notes.}

\begin{itemize}

\item All functional programs will be constructed from atomic programs through the construction rules to be presented later.
Hence all programs will contain the action of type checking for the assigned values of all elements of its input and output lists. 

\item Type checking within a program is an action that checks the type of the assigned value of a given variable.
Type checking may also include the checking of some relationship between its input variables.
For example, a type checking for integers, say $a:\mb{I}$ and $b:\mb{I}$, may include a check for value assigned equality, $a=b$, or value assigned inequality, $a<b$.
In other words a type violation error will include failure of any one of the actions of type checking, $a:\mb{I}$ and $b:\mb{I}$, and the value asigned equality or inequality 

\item Since atomic programs are constructed by an imperative programming language the notion of free parameters is difficult to avoid since atomic programs are likely to employ a number of internally defined parameters in the process of computing the output parameters from a list of input parameters.
These internally defined parameters can be regarded as free parameters that are released from storage upon execution of the program and do not appear in the output list of the atomic program.
Thus in our construction of programs by way of functional programming we will not see these free parameters.

\end{itemize}

\section{Program equivalence.}
Given a program list $P(x,y) = [P_i(x_i,y_i)]_{i=1}^n$ it is often the case that one needs to rearrange the sequential order of its subprograms.
The major constraint here is that the input of each subprogram must not depend on the output of another subprogram that follows it.
Such a rearrangement can redefine the order of the elements appearing in the input list $x$ and output list $y$ of the main program.
However, the new program obtained in this way is essentially equivalent to the original program in functionality.
Program equivalence in this sense is defined as follows. 

\begin{definition}(Program sequential equivalence.)
Two programs given by the list representations $P(x,y)=[P_i(x_i,y_i)]_{i=1}^n$ and $P^\prime(x^\prime,y^\prime)=[P_i^\prime(x_i^\prime,y_i^\prime)]_{i=1}^n$ are said to be sequential equivalent provided that $[P_i^\prime(x_i^\prime,y_i^\prime)]_{i=1}^n$ is a list permutaion of $[P_i(x_i,y_i)]_{i=1}^n$.
Sequential equivalence refers to programs where the sequential order of their subprograms has been changed without violating the I/O dependency condition.
The input/output lists of each main program $P(x,y)$ and $P^\prime(x^\prime,y^\prime)$ contain the same variable names but appear in a different order in accordence to the structure outlined in the definition of programs.
Sequential equivalence is denoted by $P(x,y) \equiv P^\prime(x^\prime,y^\prime)$.
\end{definition}

We consider another type of program equivalence, namely I/O equivalence.
Before giving the definition for I/O equivalence we need to make a distinction between common variables and constants.
For each type there may exist specific objects of that type that are of particular interest because they may appear as a fixed input parameter of a program.
For example, the type integers, $\mb{I}$, has three constants $-1,0,1$.
For higher order programs, where programs themselves serve as inputs, we may regard the empty program $[~]$ as a constant for type $\mb{P}$.
Constants of each type are introduced primarily for the purposes of defining the axioms of a given axiomatic system under consideration.

\begin{definition}(Program I/O equivalence.)
Consider two programs with the list representations $[P_i(x_i,y_i)]_{i=1}^n$ and $[P_i(x_i^\prime,y_i^\prime)]_{i=1}^n$.
For $i=1,\ldots,n$, let $m_i$ be the length of the lists $x_i$ and $x_i^\prime$ and denote by $x_{ik}$ and $x_{ik}^\prime$, respectively, the $k$th element of $x_i$ and $x_i^\prime$, respectively.
Similarly, for $i=1,\ldots,n$, let $n_i$ be the length of the lists $y_i$ and $y_i^\prime$ and denote by $y_{ik}$ and $y_{ik}^\prime$, respectively, the $k$th element of $y_i$ and $y_i^\prime$, respectively.
The two programs $[P_i(x_i,y_i)]_{i=1}^n$ and $[P_i(x_i^\prime,y_i^\prime)]_{i=1}^n$ are program I/O equivalent provided that all of the following conditions are satisfied.

\begin{itemize}

\item If $x_{ik} = x_{jl}$ then $x_{ik}^\prime = x_{jl}^\prime$, $k=1,\ldots,m_i$, $l=1,\ldots,m_j$, $i=1,\ldots,n$, $j=1,\ldots,i$.

\item If $x_{ik} = y_{jl}$ then $x_{ik}^\prime = y_{jl}^\prime$, $k=1,\ldots,m_i$, $l=1,\ldots,n_j$, $i=2,\ldots,n$, $j=1,\ldots,i-1$.

\item If $x_{ik}$ is a constant then $x_{ik}^\prime$ is also the same constant, $k=1,\ldots,m_i$, $i=1,\ldots,n$.

\end{itemize}
I/O equivalence is denoted by $[P_i(x_i,y_i)]_{i=1}^n \thicksim [P_i(x_i^\prime,y_i^\prime)]_{i=1}^n$.

\end{definition}

\section{The Sublist Rule.}

Programs are essentially strings or lists of strings with a particular structure and may serve as elements of an input/output list of a program.
A program will be said to be a higher order program if any of the elements of its input and/or output lists are themselves programs.
Here we state the program construction rules as constructs of higher order programs.
They are presented in a form that provide the link between program construction and proof construction.
As such programs can be associated with axioms, theorems and proofs. 

We use the shorthand notation
\be
\bal
& a,b,\ldots,z :\mb{P} \\
& a=P_a(x_a,y_a),b=P_b(x_b,y_b),\ldots,z=P_z(x_z,y_z) \\
\eal
\ee
Given $a:\mb{P}$ we can use the program list representation $a=[a_i]_{i=1}^n$, for some $n$.
In line with the shorthand notation $a=P_a(x_a,y_a)$, we have $a_i:\mb{P}$, $i=1,\ldots,n$, using the shorthand notation $a_i=P_{a_i}(x_{a_i},y_{a_i})$.
We may use primes to extend the range of the notation.

\begin{definition}(Computable program extension.)
A program $s:\mb{P}$ has the subtype $s:\mb{P}_{cpe}$ if it admits a decomposition $s=[p,c]$ such that if the program $p$ is computable then the program $s=[p,c]$ is also computable.
The hierachy of subtypes is $\mb{P}_{cpe}<:\mb{P}$.
The program $s=[p,c]$ is said to be a computable program extension of the program $p$.
We write $s:\mb{P}_{cpe}(p,c)$ to stress that $s$ has type $\mb{P}_{cpe}$ under the decomposition $s=[p,c]$.
\end{definition}

\begin{definition}(Irreducible computable program extension.)
A program $s:\mb{P}$ is called an irreducible computable program extension and assigned the subtype $s:\mb{P}_{icpe}$ if it admits a decomposition $s=[p,c]$ such that all of the following conditions are satisfied.

\begin{itemize}

\item $s:\mb{P}_{cpe}(p,c)$.

\item The program $p$ is irreducible in the following sense. There does not exist a program $r=[q,c]$ such that $q \subset p$ ($q$ is a strict sublist of $p$) and $r:\mb{P}_{cpe}(q,c)$.

\end{itemize}
The hierachy of subtypes is $\mb{P}_{icpe}<:\mb{P}_{cpe}<:\mb{P}$.
The programs $p$ and $c$, respectively, are said to be the premise and conclusion, respectively, of the irreducible computable program extension $s$.
We write $s:\mb{P}_{icpe}(p,c)$ to stress that $s$ has type $\mb{P}_{icpe}$ under the decomposition $s=[p,c]$.
\end{definition}

\noindent
We employ the following atomic higher order programs that are defined in the appendix.

\vspace{5mm}
\noindent
\textbf{Atomic type checking program names.}
\be
Prog,~Eqseq,~Eqio,~Sub,~Cpe
\ee

\noindent
\textbf{Atomic type asssignment program names.}
\be
Acpe
\ee

\noindent
\textbf{Atomic value assignment program names.}
\be
Conc
\ee
\vspace{5mm}

\begin{definition}(Sublist derivation.)
A sublist derivation $s:\mb{P}$ with respect to $r=[q,c]$ is an assignment $s:=[p,c]$ subject to the two conditions $q \subseteqq p$ and $r:\mb{P}_{cpe}(q,c)$.
It is constructed by the sublist derivation program $Sd([q,p,c,r],[s])$ defined by
\be
Sd([q,p,c,r],[s]) = \big [ Sub([q,p],[~]),~Cpe([q,c,r],[~]),~Conc([p,c],[s]) \big ]
\ee

\end{definition}

\vspace{5mm}
\noindent
\textbf{Proof programs.}
A program $s$ is called a proof program, or simply a proof, if it is constructed by an iteration of sublist derivations.
Consider an iteration of sublist derivations $Sd([q,p,c,r],[s])$ using the following algorithm.

\begin{itemize}

\item In the first iteration the program $p$ serves as a list of premises of the proof program.

\item For each following iteration the program $p$ is replaced by the program $s$ derived in the previous iteration.

\item In each iteration the computable program extension $r=[q,c]$ can be different.
It is selected from a collection of computable program extensions stored in memory.

\end{itemize}

\noindent
Proof programs have the further distinguishing feature in that they come with a collection of connection lists (to be defined later).
In order to present the construction rules in a concise form reference to connection lists will not be included.

\vspace{5mm}
\noindent
\textbf{Construction rules.}
The following construction rules are presented as irreducible computable program extensions of higher order constructs.
The internal square brackets deliminate the premise program from the conclusion.
(The standard list concatenation for programs apply so that the internal brackets can be removed.)
When the premise program contains only a single statement the internal square brackets are omitted.
The main inference rule is the sublist rule.
The two rules that follow it involve the acquisition of the property of a computable program extension through sequential and I/O equivalence.

The final rule states that once assigned, the property of a computable program extension is retained.
In other words, once a program has been assigned the type $\mb{P}_{cpe}$ it is stored in memory as such so that it retains that type whenever it is accessed by any following subprogram of a higher order program list. 

\vspace{5mm}
\noindent
Sublist rule.
\begin{crule}\label{slr}
\be
\bal
& \big [ Sd([q,p,c,r],[s]),~Acpe([p,c,s],[~]) \big ] \\
\eal
\ee
\end{crule}  

\noindent
Sequential equivalence of computable program extensions.
\begin{crule}\label{equiv}
\be
\bal
& \Big [ \big [ Cpe([p,c,s],[~]),~Eqseq([q,p],[~]) \big ],~Acpe([q,c,r],[~]) \Big ] \\
\eal
\ee
\end{crule}

\noindent
I/O equivalence of computable program extensions.
\begin{crule}\label{ioequiv}
\be
\bal
& \Big [ \big [ Cpe([p,c,s],[~]),~Conc([q,d],[r]),~Eqio([q,p],[~]),~Eqio([r,s],[~]) \big ],~Acpe([q,d,r],[~]) \Big ] \\
\eal
\ee
\end{crule}

\noindent
Retention of subtype assignment. 
\begin{crule}\label{rsa}
\be
\bal
& \big [ Acpe([p,c,s],[~]),~Cpe([p,c,s],[~]) \big ] \\
\eal
\ee
\end{crule}

\vspace{5mm}
\noindent
\textbf{Axioms and theorems.} Irreducible computable program extensions can be grouped into the two subtypes of axioms and theorems.

\begin{definition}(Theorem.)
A program $s:\mb{P}$ is called a theorem and assigned the subtype $s:\mb{P}_{theorem}$ if it admits a decomposition $s=[p,c]$ such that all of the following conditions are satisfied.

\begin{itemize}

\item $s:\mb{P}_{icpe}(p,c)$.

\item All theorems are extracted from programs of type $\mb{P}_{cpe}$ that have been constructed from sublist derivations.
That is, if $s=[p,c]$ is a theorem then it has been extracted from a derived program $r:\mb{P}_{cpe}(p^\prime,c)$, where the program $p^\prime$ admits the decomposition $p^\prime=[p,b]$.
The derived program $r=[p^\prime,c]=[p,b,c]$ is said to be a proof program, or simply a proof, where $p$ is the program list of the premises of the proof, $b$ is the proof body and $c$ is the conclusion of the proof.

\end{itemize}
The hierachy of subtypes is $\mb{P}_{theorem}<:\mb{P}_{icpe}<:\mb{P}_{cpe}<:\mb{P}$.
The programs $p$ and $c$, respectively, are said to be the premise and conclusion, respectively, of the theorem $s$.
We write $s:\mb{P}_{theorem}(p,c)$ to stress that $s$ has type $\mb{P}_{theorem}$ under the decomposition $s=[p,c]$.
\end{definition}

\begin{definition}(Axiom.)
A program $s:\mb{P}$ is called an axiom and assigned the subtype $s:\mb{P}_{axiom}$ if it admits a decomposition $s=[p,c]$ such that all of the following conditions are satisfied.

\begin{itemize}

\item $s:\mb{P}_{icpe}(p,c)$.

\item Axioms cannot be obtained from sublist derivations with respect to other axioms.

\end{itemize}
The hierachy of subtypes is $\mb{P}_{axiom}<:\mb{P}_{icpe}<:\mb{P}_{cpe}<:\mb{P}$.
The programs $p$ and $c$, respectively, are said to be the premise and conclusion, respectively, of the axiom $s$.
We write $s:\mb{P}_{axiom}(p,c)$ to stress that $s$ has type $\mb{P}_{axiom}$ under the decomposition $s=[p,c]$.
\end{definition}

\vspace{5mm}
\noindent
\textbf{Notes.}

\begin{itemize}

\item The above construction rules are presented as irreducible computable program extensions of higher order constructs.
They can be regarded as axioms of the axiomatic system of program constructions as proofs.
However, they do not represent the full collection of axioms from which a comprehensive analysis of program constructions as proofs can be carried out.
Such an analysis will be presented in a future paper.
The main purpose here is simply to describe the rules of program construction that largely form the basis of \emph{VPC} in its initial phase of development.

\item The sublist rule states that if $s=[p,c]$ is a sublist derivation with respect to the computable program extension $r=[q,c]$ then it follows that $s=[p,c]$ is a computable program extension of $p$.
The conclusion program of the sublist rule is a type assignment $s::\mb{P}_{cpe}(p,c)$.

\item In \emph{VPC}, the program $r=[q,c]$ of the sublist derivation $Sd([q,p,c,r],[s])$ is an axiom or theorem although the sublist rule is a more general statement in that it suffices that $r=[q,c]$ be a computable program extension.

\item During proof construction, \emph{VPC} accesses a file \emph{axiom.dat} that initially stores all of the axioms of the axiomatic system under considerations.
As proofs are completed the theorems extracted from them are also stored in \emph{axiom.dat}.

\item Axioms and theorems stored in the file \emph{axiom.dat} are automatically assigned the type $\mb{P}_{cpe}$.
Otherwise a program acquires the type $\mb{P}_{cpe}$ through the type assignment program $Acpe$.

\item In a sublist derivation $Sd([q,p,c,r],[s])$, the program $r=[q,c]$ is regarded as an axiom/theorem if it is sequential or I/O equivalent to some axiom/theorem $r^\prime=[q^\prime,c^\prime]$ stored in the file \emph{axiom.dat}.

\item A program $r=[q,d]$ may sometimes be written as an alternative concatenation $r=[q^\prime,d^\prime]$, depending on how one wishes to define the subprograms from the fully expanded program list $r=[r_i]_{i=1}^n$.
In the rule of I/O equivalence of computable program extensions we require two checks $Eqio([q,p],[~])$ and $Eqio([r,s],[~])$ to ensure that $r$ is of type $\mb{P}_{cpe}$ under the specific decomposition $r=[q,d]$.

\end{itemize}

\section{Connection List.}
For each subprogram of a program list $P(x,y)=[P_i(x_i,y_i)]_{i=1}^n$ that is obtained from a sublist derivation is constructed a connection list that records the origin of that subprogram during the program's construction.
To construct a connection list it is necessary to provide a label for each axiom and theorem of the axiomatic system under consideration.

\begin{definition}(Connection list.)
For each subprogram $P_i(x_i,y_i)$ of the program list $P(x,y)=[P_i(x_i,y_i)]_{i=1}^n$ that is obtained from a sublist derivation is generated a list that contains the labels of the axiom/theorem and premises used to obtain that subprogram.
For each such subprogram, $P_i(x_i,y_i)$, the connection list is of the form
\be
\big [ A(i),l(i,1),\ldots,l(i,k(i)) \big ]
\ee
where $1 \leq l(i,1),\ldots,l(i,k(i)) \leq i-1$ are the labels of the subprograms that make up the sublist of $P(x,y)$ that coincides with the premise program of the axiom/theorem, labelled $A(i)$, that is used to conclude $P_i(x_i,y_i)$.
Here $k(i)$ is the length of the premise program list of the axiom/theorem $A(i)$.

\end{definition}

Consider the proof program $[p,q]$, where $p=[p_i]_{i=1}^n$ is the list of premises of the proof and $q=[q_i]_{i=1}^m$ are the statements obtained by an iteration of sublist derivations.
In \emph{VPC}, proof programs are output as a vertical list with three columns.
The first column contains the statement label, the second column contains the statement itself and the third column contains the connection list.
Statements that are premises of the main proof program do not have a connection list.
The general output layout can be illustrated as follows. 

\be
\begin{array}{lll}
Label  & Statement & Connection~list \\
1      & p_1       & \\
\vdots & \vdots    & \\
n      & p_n       &  \\
n+1      & q_1     & [A(1),l(1,1),\ldots,l(1,k(1)) ]  \\
\vdots & \vdots    & \vdots \\
n+m     & q_m      & [A(m),l(m,1),\ldots,l(m,k(m)) ]  \\
\end{array}
\ee
Here $k(j)$ is the length of the premise program of the axiom/theorem $A(j)$, $j=1,\ldots,m$, and $1 \leq l(j,1),\ldots,l(j,k(j)) \leq n+j-1$, are statement labels of the sublist of the program $[p,[q_i]_{i=1}^{j-1}]$ that is sequential or I/O equivalent to the premise program of the axiom/theorem $A(j)$, $j=1,\ldots,m$, stored in \emph{axiom.dat}. 

\vspace{5mm}
\noindent
\textbf{Extraction of theorems from proofs.}
We now describe the algorithm that extracts a theorem from the proof $[p,q]$ described above.
Upon completion of a proof the final statement $q_m$ is the conclusion program of the theorem $[p,q_m]$.
Consider the list of labels
\be
c(i) = [ l(i,j) ]_{j=1}^{k(i)} , \qquad i=1,\ldots,m
\ee
obtained from the above connection lists by removing the axiom/theorem label $A(i)$. 
We construct, by iteration, a sequence of lists
\be
d(\nu) = [ \lambda(\nu,j) ]_{j=1}^{\mu(\nu)} , \qquad \nu=1,2,\ldots
\ee
where each $\lambda(\nu,j)$ is a label associated with some statement of the proof program $[p,q]$.
We may rewrite $d(\nu)$ as a partition
\be
d(\nu) = \big [ [ \lambda_p (\nu,j) ]_{j=1}^{\mu_p(\nu)}, [ \lambda_q(\nu,j) ]_{j=1}^{\mu_q(\nu)} \big ], \qquad \nu=1,2,\ldots
\ee
where $\lambda_p(\nu,j)$ are labels associated with the premise program of the proof $p$ and $\lambda_q(\nu,j)$ are labels associated with statements of $q$ obtained from sublist derivations.

For $\nu=1$ we set
\be
d(1) = c(m)
\ee
so that
\be
[ \lambda(1,j) ]_{j=1}^{\mu(1)} = [ l(m,j) ]_{j=1}^{k(m)}
\ee
and we have $\mu(1)=k(m)$ and $\lambda(1,j) = l(m,j)$, $j=1,\ldots,k(m)$.

Each list $d(\nu)$, $\nu=2,3,\dots$, consists of all labels $\lambda(\nu-1,j)$ of $d(\nu-1)$ that are associated with the premise program $p$ and the list of labels $c(\lambda(\nu-1,j))$ for labels $\lambda(\nu-1,j)$ of $d(\nu-1)$ that are associated with statements of $q$, i.e.
\be
\bal
d(\nu) = & [ \lambda(\nu,j) ]_{j=1}^{\mu(\nu)} \\
= & \big [ [ \lambda_p (\nu-1,j) ]_{j=1}^{\mu_p(\nu-1)}, [ c(\lambda_q(\nu-1,j)) ]_{j=1}^{\mu_q(\nu-1)} \big ], \qquad \nu=2,3,\ldots \\
\eal
\ee
The iteration is continued until we obtain a final list $d(\kappa)$, for some $\kappa:\mb{I}_+$, such that all of the labels of the statements contained in $q$ have been eliminated leaving only labels of the premise program $p$, i.e. $\mu_q(\kappa)=0$.
To simplify the process we may eliminate repeated labels from each list $d(\nu-1)$ before proceeding to the construction of the new list $d(\nu)$.

If by this procedure there are labels of statements in the program list $p$ that do not appear in the final list $d(\kappa)$ then those statements are redundant as premises leading to the conclusion $q_m$.
In such a case $[p,q_m]$ will not be an irreducible computable program extension and hence will not be a theorem.
The proof can be reconstructed by eliminating the redundant premises.

\vspace{5mm}
\noindent
\textbf{Options file.}  
At each step of a proof, \emph{VPC} determines all possible sublist derivations that can be obtained from the main program with respect to the axioms and theorems that are currently stored in the file \emph{axiom.dat}.
These sublist derivations are listed in an options file which the user may consult and select the desired conclusion program to generate a new statement in the main program list.
The process is repeated until the proof is completed.

Each option in the options file includes the axiom/theorem label and the associated labels of the subprograms that make up the sublist of the current proof program that coincides with the premise program of the axiom/theorem stored in \emph{axiom.dat}.
Crucial to this search and matching procedure is program sequential and I/O equivalence.
Extensions of sublists of the proof program acquire the property of axioms and theorems from the construction rules of sequential and I/O equivalence of computable program extensions.

Extractions of sublists from the current proof program based on a raw search of all possible permutations followed by an I/O equivalence matching algorithm can be computatonally expensive.
\emph{VPC} employs special techniques that speed up this process by detecting and eliminating unsuccesful matches before a complete sublist extraction and I/O equivalence check is required.
This significantly reduces the computations making the enumeration of all possible sublist derivations quite manageable.

\section{Disjunctions.}

Disjunctions play an important role in the expressiveness of formal statements in standard theories of logic.
For well structured computer programs, disjunctions play a more elementray role in that they effectively split a main program into several parallel programs, each of which is associated with an operand of the disjunction.
Sublist derivations can be applied independently to each parallel program to form a system of simultaneous program constructions.
If sublist derivations of the operand programs lead to a common conclusion then that conclusion can be contracted back onto the main program.  

Let $a:\mb{P}$ and $b:\mb{P}$ be programs that are the operands of the disjunction program $q=a~|~b$, (to be read $a$ or $b$).
Disjunction splitting and contraction follow from the right and left distributivity rules
\be
[p,q]=[p,a~|~b] = [p,a]~|~[p,b]
\ee
\be
[q,p] = [a~|~b,p] = [a,p]~|~[b,p]
\ee
Note that we require two independent rules of right and left distributivity because subprogram sequential reordering is constrained by the I/O dependency condition.

Consider a proof program under construction that takes the form $[p,a~|~b]$.
We may wish to perform a disjunction splitting that follows from the right distributivity rule to obtain $[p,a]~|~[p,b]$.
Suppose that we have independently applied sublist derivations to the operand programs $[p,a]$ and $[p,b]$ to obtain a common conclusion $[[p,a],c]$ and $[[p,b],c]$.
We can now apply the left and then right distributivity rules to obtain
\be
\big [ [p,a],c \big ]~|~\big [ [p,b],c \big ] = \big [ [p,a]~|~[p,b],c \big ] = [ p,a~|~b,c] = [p,q,c]
\ee
Note that the disjunction program $a~|~b$ is not a program list but rather an element program of the program list $[ p,a~|~b,c]$.

There is also the possibility that while the main proof program $[p,a~|~b]$ is computable, one of the operand programs $[p,a]$ or $[p,b]$ may not be computable.
Operand programs that are not computable can simply be discarded.
Suppose for instance that $[p,b]$ is not computable and we have by a sublist derivation $[[p,a],c]$.
In such a case we simply discard the operand program $[p,b]$ and contract the conclusion $c$ of the operand program $[[p,a],c]$ onto the main program.

For this to work we must have a procedure from which one can determine whether a given program is not computable.
This requires additional axioms from which it is possible to employ sublist derivations that lead to a conclusion of noncomputability.
This is a subtopic that requires a fairly detailed discussion involving higher order axioms and will be postponed for a future paper.

All of the above may be readily extended to cases where there exists more than two operands.
Many tests have already been performed in \emph{VPC} using various procedures for dealing with disjunctions, each procedure having its own advantages and disadvantages.
As with methods for determining noncomputability, a detailed discussion of disjunctions will be postoned for a future paper.
Here, whenever a disjunction may be relevant, we simply consider each operand program as a separate proof program.
The version of \emph{VPC}, as presented here, will simply lack the feature of contracting a common conclusion to a main proof program containing a disjunction.  

\section{Arithmetic on $\mb{I}$.}

We work with objects of type $\mb{I}$ that can be assigned any of the integer values 
\be
0, \pm 1,\ldots, \pm N,
\ee
where $N$ is the maximum positive integer that is a machine dependent parameter.
The objective here is to construct an axiomatic system for arithmetic on $\mb{I}$ that reflects the practical issues of machine arithmetic where exact integer solutions are required.
Applications are found in many integer programming problems that include static and dynamic network flows and the solution of systems of Diophantine equations (see for example \cite{bert}, \cite{orli}, \cite{waer}, \cite{boro}, \cite{green}). 

A major hurdle when working on $\mb{I}$ is the lack of closure of the elementary arithmetic operations of addition and multiplication.
Here we shall take a less abstract approach to the standard theory of commutative rings by introducing rules that address the operations of machine arithmetic that lend themselves to a more practical approach to the verification of computability.
The results that will be presented in a later section serve as a first step towards a more detailed analysis of operations and manipulations of matrices on $\mb{I}$.
Integer matrix operations will be postponed for a future paper.

We could also work with a finite subset of the rationals $\mb{X} = \epsilon \mb{I}$, where $0<\epsilon << 1$ is also a machine specific parameter.
The finite subset of the rationals $\mb{X}$ has a fixed size resolution so that arithmetic on $\mb{X}$, as defined here, differs from floating point arithmetic. 
Because of this many results of arithmetic on $\mb{I}$ can be directly applied to $\mb{X}$, although there are significant departures that would warrant a separate analysis.

For arithmetic on $\mb{I}$ we employ the following atomic integer programs that are defined in the appendix.

\vspace{5mm}
\noindent
\textbf{Atomic type checking program names.} 
\be
Int,~Lt,~Eq,~Neq
\ee

\noindent
\textbf{Atomic value assignment program names.}
\be
Aid,~Add,~Mult,~Div
\ee

There are three constants, $-1,0$ and $1$, that will sometimes appear in the input lists of some programs.
Recall that elements of an input and output list of a program are alphanumeric variable names that serve as placeholders for assigned values of the type specified within the program.
We depart from this convention slightly such that whenever the constants $-1,0$ and $1$ appear in an input list they are to be regarded as variable names that are placeholders that accept only the corresponding integer value.

Axioms are labelled with an uppper case $A$ followed by a number.
These axioms are stored in a file $axiom.dat$ that is accessed by \emph{VPC} during proof construction.
In the following axioms the internal square brackets deliminate the premise program from the conclusion.
(The standard list concatenation for programs apply so that the internal brackets can be removed.)
When there is only a single premise the internal square brackets are omitted.

\vspace{5mm}
\noindent
\textbf{Identity axioms for type $\mb{I}$ I/O lists.}

\begin{axm}\label{idtype1}
\be
\big [ P(x,y),~Int([x_i],[~]) \big ] , \qquad x_i \in x
\ee
\end{axm}
\begin{axm}\label{idtype2}
\be
\big [ P(x,y),~Int([y_i],[~]) \big ] , \qquad y_i \in y
\ee
\end{axm}

\noindent
Here $x$ and $y$ are lists and $P$ is a generic integer program name. 

\vspace{5mm}
\noindent
\textbf{Axiom of substitution.}
To present the axiom of substitution in a more general form we introduce the nonatomic list equality program $Eqlst$.
For two integer lists $u=[u_i]_{i=1}^n$ and $v=[v_i]_{i=1}^n$, $n:\mb{I}_+$, we define the program $Eqlst$ as
\be
\bal
Eqlst([u_1,v_1,u_2,v_2,\ldots,u_n,v_n],[~]) = & \Big [ Eq([u_1,v_1],[~]),~Eq([u_2,v_2],[~]),\ldots,~Eq([u_n,v_n],[~]) \Big ] \\
= & \Big [ Eq([u_i,v_i],[~]) \Big ]_{i=1}^n \\
\eal
\ee
For the case $n=1$ the list equality program $Eqlst$ reduces to the atomic program $Eq$, i.e.
\be
Eqlst([u_1,v_1],[~]) = Eq([u_1,v_1],[~])
\ee
The first part of the axiom of substitution is an existence axiom.

\begin{axm}\label{subst1}
\be
\Big [ \big [ P(x,y),~Eq([x_i,a],[~]) \big ],~P(x^\prime,y^\prime) \Big ], \qquad x_i \in x ,~x^\prime = x|_{x_i \to a}
\ee
\end{axm}
\noindent
Here $x$, $x^\prime$, $y$ and $y^\prime$ are lists and $P$ is a generic integer program name.
Both $y$ and $y^\prime$ may be the empty list.
The second part of the axiom of substitution is applicable when $y$ and $y^\prime$ are not empty lists and establishes their equality.
\begin{axm}\label{subst2}
\be
\Big [ \big [ P(x,y),~Eq([x_i,a],[~]),~P(x^\prime,y^\prime) \big ],~Eqlst([y_1^\prime,y_1,y_2^\prime,y_2,\ldots,y_n^\prime,y_n],[~]) \Big ], \qquad x_i \in x ,~x^\prime = x|_{x_i \to a}
\ee 
\end{axm}

\noindent
Here $y$ and $y^\prime$ are represented as lists of length $n:\mb{I}_+$.
For the case $n=1$ the conclusion program reduces to $Eq([y_1^\prime,y_1],[~])$. 

\vspace{5mm}
\noindent
\textbf{Equality axioms.}

\noindent
Reflexivity.
\begin{axm}\label{eql1}
\be
\big [ Int([a],[~]),~Eq([a,a],[~]) \big ]
\ee
\end{axm}

\noindent
Symmetry.
\begin{axm}\label{eql2}
\be
\big [ Eq([a,b],[~]),~Eq([b,a],[~]) \big ]
\ee
\end{axm}

\noindent
Transitivity of equality states that
\be
\Big [ \big [ Eq([a,b],[~]),~Eq([b,c],[~]) \big ],~Eq([a,c],[~]) \Big ]
\ee
We do not include this as an axiom because it can be derived from the first part of the axiom of substitution.

\vspace{5mm}
\noindent
\textbf{Identity assignment axiom.}

\begin{axm}\label{as}
\be
\big [ Aid([a],[b]),~Eq([b,a],[~]) \big ]
\ee
\end{axm}

\vspace{5mm}
\noindent
\section{Axioms of arithmetic on $\mb{I}$.}

\vspace{5mm}
\noindent
\textbf{Axioms of addition and multiplication.}

\noindent
Commutivity of addition.

\begin{axm}\label{addcom1}
\be
\big [ Add([a,b],[c]),~Add([b,a],[d]) \big ]
\ee
\end{axm}

\begin{axm}\label{addcom2}
\be
\Big [ \big [ Add([a,b],[c]),~Add([b,a],[d]) \big ],~Eq([d,c],[~]) \Big ]
\ee
\end{axm}

\noindent
Associativity of addition.

\begin{axm}\label{addassoc1}
\be
\Big [ \big [ Add([a,b],[d]),~Add([d,c],[x]),~Add([b,c],[e]) \big ],~Add([a,e],[y]) \Big ]
\ee
\end{axm}

\begin{axm}\label{addassoc2}
\be
\bal
& \Big [ \big [ Add([a,b],[d]),~Add([d,c],[x]),~Add([b,c],[e]),~Add([a,e],[y]) \big ], Eq([y,x],[~]) \Big ] \\
\eal
\ee
\end{axm}

\noindent
Addition by zero.

\begin{axm}\label{addz1}
\be
\big [ Int([a],[~]),~Add([a,0],[b]) \big ]
\ee
\end{axm}
\begin{axm}\label{addz2}
\be
\big [ Add([a,0],[b]),~Eq([b,a],[~]) \big ]
\ee
\end{axm}

\noindent
Additive inverse.

\begin{axm}\label{multinv1}
\be
\big [ Int([a],[~]),~Mult([-1,a],[b]) \big ]
\ee
\end{axm}
\begin{axm}\label{multinv2}
\be
\big [ Mult([-1,a],[b]),~Add([a,b],[d]) \big ]
\ee
\end{axm}
\begin{axm}\label{multinv3}
\be
\Big [ \big [ Mult([-1,a],[b]),~Add([a,b],[d]) \big ],~Eq([d,0],[~]) \Big ]
\ee
\end{axm}

\noindent
Commutivity of multiplication.

\begin{axm}\label{multcomm1}
\be
\big [ Mult([a,b],[c]),~Mult([b,a],[d]) \big ]
\ee
\end{axm}
\begin{axm}\label{multcomm2}
\be
\Big [ \big [ Mult([a,b],[c]),~Mult([b,a],[d]) \big ],~Eq([d,c],[~]) \Big ]
\ee
\end{axm}

\noindent
Associativity of multiplication.

\begin{axm}\label{multassoc1}
\be
\Big [ \big [ Mult([a,b],[d]),~Mult([d,c],[x]),~Mult([b,c],e] \big ],~Mult([a,e],[y]) \Big ]
\ee
\end{axm}
\begin{axm}\label{multassoc2}
\be
\Big [ \big [ Mult([a,b],[d]),~Mult([d,c],[x]),~Mult([b,c],e],~Mult([a,e],[y]) \big ],~Eq([y,x],[~]) \Big ]
\ee
\end{axm}

\noindent
Multiplication by unity.

\begin{axm}\label{multun1}
\be
\big [ Int([a],[~]),~Mult([1,a],[b]) \big ]
\ee
\end{axm}
\begin{axm}\label{multun2}
\be
\big [ Mult([1,a],[b]),~Eq([b,a],[~]) \big ]
\ee
\end{axm}

\noindent
Distributive law.

\begin{axm}\label{dist1}
\be
\Big [ \big [ Add([b,c],[d]),~Mult([a,d],[x]),~Mult([a,b],[u]),~Mult([a,c],[v]) \big ],~Add([u,v],[y]) \Big ]
\ee
\end{axm}
\begin{axm}\label{dist2}
\be
\Big [ \big [ Mult([a,b],[u]),~Mult([a,c],[v]),~Add([u,v],[y]),~Add([b,c],[d]) \big ],~Mult([a,d],[x]) \Big ]
\ee
\end{axm}
\begin{axm}\label{dist3}
\be
\bal
& \Big [ \big [ Add([b,c],[d]),~Mult([a,d],[x]),~Mult([a,b],[u]),~Mult([a,c],[v]),~Add([u,v],[y]) \big ],~Eq([y,x],[~]) \Big ] \\
\eal
\ee
\end{axm}

\vspace{5mm}
\noindent
\textbf{Order axioms.}

\begin{axm}\label{ordadd}
\be
\Big [ \big [ Lt([a,b],[~]),~Add([a,c],[x]),~Add([b,c],[y]) \big ],~Lt([x,y],[~]) \Big ]
\ee
\end{axm}
\begin{axm}\label{ordmult1}
\be
\Big [ \big [ Lt([a,b],[~]),~Lt([0,c],[~]),~Mult([a,c],[x]),~Mult([b,c],[y]) \big ],~Lt([x,y],[~]) \Big ]
\ee
\end{axm}
\begin{axm}\label{ordmult2}
\be
\Big [ \big [ Lt([a,b],[~]),~Lt([c,0],[~]),~Mult([a,c],[x]),~Mult([b,c],[y]) \big ],~Lt([y,x],[~]) \Big ]
\ee
\end{axm}

\noindent
Transitivity.

\begin{axm}\label{ordtrans}
\be
\Big [ \big [ Lt([a,b],[~]),~Lt([b,c],[~]) \big ],~Lt([a,c],[~]) \Big ]
\ee
\end{axm}

\vspace{5mm}
\noindent
\textbf{Divisor.}

\begin{axm}\label{div1}
\be
\Big [ \big [ Neq([a,0],[~]),~Mult([a,b],[c]) \big ],~Div([c,a],[d]) \Big ]
\ee
\end{axm}
\begin{axm}\label{div2}
\be
\Big [ \big [ Mult([a,b],[c]),~Div([c,a],[d]) \big ],~Eq([d,b],[~]) \Big ]
\ee
\end{axm}

\vspace{5mm}
\noindent
\textbf{Notes.}

\begin{itemize}

\item The above axioms for addition and multiplication differ from the axioms of fields and commutative rings because we do not have closure,
i.e. $Add([a,b],[c])$ and $Mult([a,b],[c])$ do not necessarily follow from $Int([a],[~])$ and $Int([b],[~])$.
Because of this the above axioms for addition and multiplication are split into one or more existence parts followed by an equality axiom.
Any occurence of statements involving $Add$ and $Mult$ in a program list must either have been inferred from the axioms or have simply been inserted as premises.

\item The additive inverse is defined explicitly as a multiplication $Mult([-1,a],[b])$.

\item Associativity of addition requires an existence axiom $A10$ followed by an equality axiom $A11$.
The existence part is necessary because $y=a+(b+c)$ does not follow from $x=(a+b)+c$.
As an example set $a=-N,~b=N,~c=1$.
We have $d=a+b=0:\mb{I}$ and hence $x=(-N+N)+1=1:\mb{I}$ but $e=b+c=N+1$, and hence $y=a+(b+c)$, is not of type $\mb{I}$. 
In order that $y=a+(b+c):\mb{I}$ we must first gaurantee that $e=b+c:\mb{I}$.

\item Associativity of multiplication requires an existence axiom $A19$ followed by an equality axiom $A20$.
The existence part is necessary because $y=a(bc)$ does not follow from $x=(ab)c$.
As an example set $a=0,~b=N,~c=2$.
We have $d=ab=0:\mb{I}$ and hence $x=(0)2=0:\mb{I}$ but $e=bc=N.2$ which is not of type $\mb{I}$. 
In order that $y=a(bc):\mb{I}$ we must first gaurantee that $e=bc:\mb{I}$.

\item The axiom of distributivity has two independent existence parts, $A23$ and $A24$, followed by an equality axiom, $A25$.
The existence part $A23$ is necessary because $y=ab+ac$ does not follow from $x=a(b+c)$.
As an example set $a=N,~b=N,~c=-N$.
We have $d=b+c=0:\mb{I}$ and hence $x=N(-N+N)=0:\mb{I}$ but neither $u=ab$ and $v=ac$, and hence $y=u+v$, are of type $\mb{I}$. 
In order that $y=ab+ac:\mb{I}$ we must first gaurantee that $u$ and $v$ are of type $\mb{I}$.
The existence part $A24$ is necessary because $x=a(b+c)$ does not follow from $y=ab+ac$.
As an example set $a=0,~b=N,~c=N$.
We have $u=ab=0:\mb{I}$ and $v=ac=0:\mb{I}$ and hence their sum $y=0:\mb{I}$.
But $d=b+c=N+N$ is not of type $\mb{I}$.
In order that $x=a(b+c):\mb{I}$ we must first gaurantee that $d=b+c$ is of type $\mb{I}$.

\item In the theory of fields and commutative rings the statements similar to $A27$ and $A28$ are theorems that can be obtained from the single axiom that states that if $x>0$ and $y>0$ then $xy>0$.
For arithmetic on $\mb{I}$ this axiom cannot be used to prove $A27$ and $A28$ since additional premises are necessary.
The additional premises would render the axiom as unnecessarily restrictive.
For this reason $A27$ and $A28$ are presented as axioms.

\end{itemize}

\section{Basic identities and inequalities on $\mb{I}$.}

Derivations of proofs in arithmetic on $\mb{I}$ can sometimes be much lengthier than their counterparts in field and ring theory.
The main difficulty arises from the absence of closure of addition and multiplication.
As a consequence many proofs are actually dedicated to the establishment of existence.

At each step of a proof construction, \emph{VPC} accesses the data file \emph{axiom.dat} that initially stores all of the axioms of the axiomatic system under consideration.
As proofs are completed the theorems extracted from them are automatically appended to the file \emph{axiom.dat}.
All axioms and theorems will be provided with a label that are stored in \emph{axiom.dat}.
There are two label types: $A$ followed by a number is an axiom; $T$ followed by a number is a theorem.

The proofs presented below were generated manually.
At each step of a proof, \emph{VPC} determines all possible sublist derivations that can be obtained from the current proof program.
These are listed in an options file that the user can consult to select a desired conclusion program.
Each option includes the axiom/theorem label and the associated labels of the element programs of the sublist of the current proof program that coincides with the premise program of the axiom/theorem stored in \emph{axiom.dat}.
The user then selects the desired option (conclusion program) to generate a new statement in the program list.
The process is repeated until the proof is completed.
Crucial to the matching procedure of sublists of the proof program with premise programs of the axioms/theorems stored in \emph{axiom.dat} is program sequential and I/O equivalence.

Theorems are presented as horizontal lists while proof programs are presented as vertical lists. 
The first entry of each line of a proof is the program label (equivalent to the element number) followed by the statement.
Following the statement is the connection list in which the first entry is the axiom/theorem label and the remaining labels associated with the premises used to generate the current statement from a sublist derivation.
The absence of a connection list means that the statement is a premise of the main proof program. 
In the proofs given below, \emph{VPC} provides an optional additional entry that attempts to assist the user in keeping track of the relationships between variables.

Many theorems that are presented below come in pairs, the first part establishing existence and the second part establishing the equality.
The proofs are presented for demonstration purposes only and are not meant to represent the most efficient or elegant proof of the given theorem.
We start with some elementary theorems related to the identities of algebra ($T1$-$T13$) and then proceed to inequalities ($T14$-$T17$).
The listings are imported directly from an output file generated by \emph{VPC}.

\begin{lstlisting}

Theorem T1.
[[Add([a,b],[c]), Mult([-1,b],[d])], Add([c,d],[m])]

Proof.
  1 Add([a,b],[c])                          c=(a+b)
  2 Mult([-1,b],[d])                        d=(-1*b)
  3 Add([b,d],[e])      [A15,2]             e=(b+d)=(b+(-1*b))
  4 Add([d,b],[f])      [A8,3]              f=(d+b)=((-1*b)+b)
  5 Add([b,a],[g])      [A8,1]              g=(b+a)
  6 Eq([e,0],[])        [A16,2,3]           e=0
  7 Eq([e,f],[])        [A9,4,3]            e=f
  8 Eq([0,f],[])        [A3,7,6]            0=f
  9 Eq([g,c],[])        [A9,1,5]            g=c
 10 Int([a],[])         [A1,1]              a:I
 11 Add([a,0],[h])      [A12,10]            h=(a+0)
 12 Add([0,a],[i])      [A8,11]             i=(0+a)
 13 Add([f,a],[j])      [A3,12,8]           j=(f+a)=(((-1*b)+b)+a)
 14 Add([d,g],[k])      [A10,4,13,5]        k=(d+g)=((-1*b)+(b+a))
 15 Add([d,c],[l])      [A3,14,9]           l=(d+c)=((-1*b)+(a+b))
 16 Add([c,d],[m])      [A8,15]             m=(c+d)=((a+b)+(-1*b))

Theorem T2.
[[Add([a,b],[c]), Mult([-1,b],[d]), Add([c,d],[m])], Eq([m,a],[])]

Proof.
  1 Add([a,b],[c])                          c=(a+b)
  2 Mult([-1,b],[d])                        d=(-1*b)
  3 Add([c,d],[m])                          m=(c+d)=((a+b)+(-1*b))
  4 Add([b,d],[e])      [A15,2]             e=(b+d)=(b+(-1*b))
  5 Add([d,b],[f])      [A8,4]              f=(d+b)=((-1*b)+b)
  6 Add([b,a],[g])      [A8,1]              g=(b+a)
  7 Eq([e,0],[])        [A16,2,4]           e=0
  8 Eq([e,f],[])        [A9,5,4]            e=f
  9 Eq([0,f],[])        [A3,8,7]            0=f
 10 Eq([g,c],[])        [A9,1,6]            g=c
 11 Int([a],[])         [A1,1]              a:I
 12 Add([a,0],[h])      [A12,11]            h=(a+0)
 13 Add([0,a],[i])      [A8,12]             i=(0+a)
 14 Add([f,a],[j])      [A3,13,9]           j=(f+a)=(((-1*b)+b)+a)
 15 Add([d,g],[k])      [A10,5,14,6]        k=(d+g)=((-1*b)+(b+a))
 16 Add([d,c],[l])      [A3,15,10]          l=(d+c)=((-1*b)+(a+b))
 17 Eq([m,l],[])        [A9,16,3]           m=l
 18 Eq([l,k],[])        [A4,15,10,16]       l=k
 19 Eq([k,j],[])        [A11,5,14,6,15]     k=j
 20 Eq([l,j],[])        [A3,18,19]          l=j
 21 Eq([m,j],[])        [A3,17,20]          m=j
 22 Eq([j,i],[])        [A4,13,9,14]        j=i
 23 Eq([m,i],[])        [A3,21,22]          m=i
 24 Eq([i,h],[])        [A9,12,13]          i=h
 25 Eq([m,h],[])        [A3,23,24]          m=h
 26 Eq([h,a],[])        [A13,12]            h=a
 27 Eq([m,a],[])        [A3,25,26]          m=a

Theorem T3.
[[Add([a,b],[c]), Add([a,d],[e]), Eq([c,e],[])], Eq([b,d],[])]

Proof.
  1 Add([a,b],[c])                          c=(a+b)
  2 Add([a,d],[e])                          e=(a+d)
  3 Eq([c,e],[])                            c=e
  4 Add([b,a],[f])      [A8,1]              f=(b+a)
  5 Add([d,a],[g])      [A8,2]              g=(d+a)
  6 Eq([f,c],[])        [A9,1,4]            f=c
  7 Eq([g,e],[])        [A9,2,5]            g=e
  8 Int([a],[])         [A1,1]              a:I
  9 Mult([-1,a],[h])    [A14,8]             h=(-1*a)
 10 Add([f,h],[i])      [T1,4,9]            i=(f+h)=((b+a)+(-1*a))
 11 Add([g,h],[j])      [T1,5,9]            j=(g+h)=((d+a)+(-1*a))
 12 Eq([i,b],[])        [T2,4,9,10]         i=b
 13 Eq([j,d],[])        [T2,5,9,11]         j=d
 14 Add([c,h],[k])      [A3,10,6]           k=(c+h)=((a+b)+(-1*a))
 15 Add([e,h],[l])      [A3,11,7]           l=(e+h)=((a+d)+(-1*a))
 16 Eq([k,i],[])        [A4,10,6,14]        k=i
 17 Eq([l,j],[])        [A4,11,7,15]        l=j
 18 Eq([l,k],[])        [A4,14,3,15]        l=k
 19 Eq([k,b],[])        [A3,16,12]          k=b
 20 Eq([l,d],[])        [A3,17,13]          l=d
 21 Eq([k,l],[])        [A6,18]             k=l
 22 Eq([b,l],[])        [A3,21,19]          b=l
 23 Eq([b,d],[])        [A3,22,20]          b=d

Theorem T4.
[[Mult([a,b],[c]), Mult([a,d],[e]), Eq([c,e],[]), Neq([a,0],[])], Eq([b,d],[])]

Proof.
  1 Mult([a,b],[c])                         c=(a*b)
  2 Mult([a,d],[e])                         e=(a*d)
  3 Eq([c,e],[])                            c=e
  4 Neq([a,0],[])                           a/=0
  5 Div([c,a],[f])      [A30,4,1]           f=(c/a)=((a*b)/a)
  6 Div([e,a],[g])      [A3,5,3]            g=(e/a)=((a*d)/a)
  7 Eq([f,b],[])        [A31,1,5]           f=b
  8 Eq([g,d],[])        [A31,2,6]           g=d
  9 Eq([g,f],[])        [A4,5,3,6]          g=f
 10 Eq([g,b],[])        [A3,9,7]            g=b
 11 Eq([b,d],[])        [A3,8,10]           b=d

Theorem T5.
[[Int([a],[])], Mult([0,a],[o])]

Proof.
  1 Int([a],[])                             a:I
  2 Mult([1,a],[b])     [A21,1]             b=(1*a)
  3 Eq([b,a],[])        [A22,2]             b=a
  4 Mult([a,1],[c])     [A17,2]             c=(a*1)
  5 Eq([c,b],[])        [A18,2,4]           c=b
  6 Eq([c,a],[])        [A3,5,3]            c=a
  7 Mult([-1,a],[d])    [A14,1]             d=(-1*a)
  8 Add([a,d],[e])      [A15,7]             e=(a+d)=(a+(-1*a))
  9 Mult([a,-1],[f])    [A17,7]             f=(a*-1)
 10 Eq([f,d],[])        [A18,7,9]           f=d
 11 Int([1],[])         [A1,2]              1:I
 12 Mult([-1,1],[g])    [A14,11]            g=(-1*1)
 13 Add([1,g],[h])      [A15,12]            h=(1+g)=(1+(-1*1))
 14 Eq([h,0],[])        [A16,12,13]         h=0
 15 Mult([1,-1],[i])    [A17,12]            i=(1*-1)
 16 Eq([i,-1],[])       [A22,15]            i=-1
 17 Eq([g,i],[])        [A18,15,12]         g=i
 18 Eq([g,-1],[])       [A3,17,16]          g=-1
 19 Add([1,-1],[j])     [A3,13,18]          j=(1+-1)
 20 Eq([j,h],[])        [A4,13,18,19]       j=h
 21 Eq([j,0],[])        [A3,20,14]          j=0
 22 Eq([a,c],[])        [A6,6]              a=c
 23 Add([c,d],[k])      [A3,8,22]           k=(c+d)=((a*1)+(-1*a))
 24 Eq([d,f],[])        [A6,10]             d=f
 25 Add([c,f],[l])      [A3,23,24]          l=(c+f)=((a*1)+(a*-1))
 26 Mult([a,j],[m])     [A24,4,9,25,19]     m=(a*j)=(a*(1+-1))
 27 Mult([j,a],[n])     [A17,26]            n=(j*a)=((1+-1)*a)
 28 Mult([0,a],[o])     [A3,27,21]          o=(0*a)

Theorem T6.
[[Mult([0,a],[b])], Eq([b,0],[])]

Proof.
  1 Mult([0,a],[b])                         b=(0*a)
  2 Int([a],[])         [A1,1]              a:I
  3 Mult([1,a],[c])     [A21,2]             c=(1*a)
  4 Eq([c,a],[])        [A22,3]             c=a
  5 Mult([a,1],[d])     [A17,3]             d=(a*1)
  6 Eq([d,c],[])        [A18,3,5]           d=c
  7 Mult([-1,a],[e])    [A14,2]             e=(-1*a)
  8 Mult([a,-1],[f])    [A17,7]             f=(a*-1)
  9 Int([-1],[])        [A1,7]              -1:I
 10 Mult([1,-1],[g])    [A21,9]             g=(1*-1)
 11 Eq([g,-1],[])       [A22,10]            g=-1
 12 Mult([-1,1],[h])    [A17,10]            h=(-1*1)
 13 Eq([h,g],[])        [A18,10,12]         h=g
 14 Eq([h,-1],[])       [A3,13,11]          h=-1
 15 Add([1,h],[i])      [A15,12]            i=(1+h)=(1+(-1*1))
 16 Eq([i,0],[])        [A16,12,15]         i=0
 17 Add([1,-1],[j])     [A3,15,14]          j=(1+-1)
 18 Eq([j,i],[])        [A4,15,14,17]       j=i
 19 Eq([j,0],[])        [A3,18,16]          j=0
 20 Eq([0,j],[])        [A6,19]             0=j
 21 Mult([j,a],[k])     [A3,1,20]           k=(j*a)=((1+-1)*a)
 22 Eq([k,b],[])        [A4,1,20,21]        k=b
 23 Mult([a,j],[l])     [A17,21]            l=(a*j)=(a*(1+-1))
 24 Eq([l,k],[])        [A18,21,23]         l=k
 25 Eq([l,b],[])        [A3,24,22]          l=b
 26 Add([d,f],[m])      [A23,17,23,5,8]     m=(d+f)=((a*1)+(a*-1))
 27 Eq([m,l],[])        [A25,17,23,5,8,26]  m=l
 28 Eq([m,b],[])        [A3,27,25]          m=b
 29 Eq([f,e],[])        [A18,7,8]           f=e
 30 Add([d,e],[n])      [A3,26,29]          n=(d+e)=((a*1)+(-1*a))
 31 Eq([n,m],[])        [A4,26,29,30]       n=m
 32 Eq([n,b],[])        [A3,31,28]          n=b
 33 Add([c,e],[o])      [A3,30,6]           o=(c+e)=((1*a)+(-1*a))
 34 Eq([o,n],[])        [A4,30,6,33]        o=n
 35 Eq([o,b],[])        [A3,34,32]          o=b
 36 Add([a,e],[p])      [A3,33,4]           p=(a+e)=(a+(-1*a))
 37 Eq([p,o],[])        [A4,33,4,36]        p=o
 38 Eq([p,0],[])        [A16,7,36]          p=0
 39 Eq([b,o],[])        [A6,35]             b=o
 40 Eq([o,0],[])        [A3,38,37]          o=0
 41 Eq([b,0],[])        [A3,39,40]          b=0

Theorem T7.
[[Mult([-1,a],[b]), Mult([-1,b],[c])], Eq([c,a],[])]

Proof.
  1 Mult([-1,a],[b])                        b=(-1*a)
  2 Mult([-1,b],[c])                        c=(-1*b)=(-1*(-1*a))
  3 Add([a,b],[d])      [A15,1]             d=(a+b)=(a+(-1*a))
  4 Add([b,c],[e])      [A15,2]             e=(b+c)=((-1*a)+(-1*(-1*a)))
  5 Eq([d,0],[])        [A16,1,3]           d=0
  6 Eq([e,0],[])        [A16,2,4]           e=0
  7 Eq([0,e],[])        [A6,6]              0=e
  8 Eq([d,e],[])        [A3,5,7]            d=e
  9 Add([b,a],[f])      [A8,3]              f=(b+a)=((-1*a)+a)
 10 Eq([f,d],[])        [A9,3,9]            f=d
 11 Eq([f,e],[])        [A3,10,8]           f=e
 12 Eq([a,c],[])        [T3,9,4,11]         a=c
 13 Eq([c,a],[])        [A6,12]             c=a

Theorem T8.
[[Mult([a,b],[c]), Mult([-1,b],[d])], Mult([a,d],[i])]

Proof.
  1 Mult([a,b],[c])                         c=(a*b)
  2 Mult([-1,b],[d])                        d=(-1*b)
  3 Int([c],[])         [A2,1]              c:I
  4 Mult([-1,c],[e])    [A14,3]             e=(-1*c)=(-1*(a*b))
  5 Mult([b,-1],[f])    [A17,2]             f=(b*-1)
  6 Mult([c,-1],[g])    [A17,4]             g=(c*-1)=((a*b)*-1)
  7 Eq([f,d],[])        [A18,2,5]           f=d
  8 Mult([a,f],[h])     [A19,1,6,5]         h=(a*f)=(a*(b*-1))
  9 Mult([a,d],[i])     [A3,8,7]            i=(a*d)=(a*(-1*b))

Theorem T9.
[[Mult([a,b],[c]), Mult([-1,b],[d]), Mult([a,d],[i]), Mult([-1,c],[e])], Eq([i,e],[])]

Proof.
  1 Mult([a,b],[c])                         c=(a*b)
  2 Mult([-1,b],[d])                        d=(-1*b)
  3 Mult([a,d],[i])                         i=(a*d)=(a*(-1*b))
  4 Mult([-1,c],[e])                        e=(-1*c)=(-1*(a*b))
  5 Mult([b,-1],[f])    [A17,2]             f=(b*-1)
  6 Mult([c,-1],[g])    [A17,4]             g=(c*-1)=((a*b)*-1)
  7 Eq([f,d],[])        [A18,2,5]           f=d
  8 Eq([g,e],[])        [A18,4,6]           g=e
  9 Mult([a,f],[h])     [A19,1,6,5]         h=(a*f)=(a*(b*-1))
 10 Eq([h,g],[])        [A20,1,6,5,9]       h=g
 11 Eq([i,h],[])        [A4,9,7,3]          i=h
 12 Eq([i,g],[])        [A3,11,10]          i=g
 13 Eq([i,e],[])        [A3,12,8]           i=e

Theorem T10.
[[Mult([a,b],[c]), Mult([-1,a],[d])], Mult([d,b],[g])]

Proof.
  1 Mult([a,b],[c])                         c=(a*b)
  2 Mult([-1,a],[d])                        d=(-1*a)
  3 Mult([b,a],[e])     [A17,1]             e=(b*a)
  4 Mult([b,d],[f])     [T8,3,2]            f=(b*d)=(b*(-1*a))
  5 Mult([d,b],[g])     [A17,4]             g=(d*b)=((-1*a)*b)

Theorem T11.
[[Mult([a,b],[c]), Mult([-1,a],[d]), Mult([d,b],[g]), Mult([-1,c],[h])], Eq([g,h],[])]

Proof.
  1 Mult([a,b],[c])                         c=(a*b)
  2 Mult([-1,a],[d])                        d=(-1*a)
  3 Mult([d,b],[g])                         g=(d*b)=((-1*a)*b)
  4 Mult([-1,c],[h])                        h=(-1*c)=(-1*(a*b))
  5 Eq([h,g],[])        [A20,2,3,1,4]       h=g
  6 Eq([g,h],[])        [A6,5]              g=h

Theorem T12.
[[Mult([a,b],[c]), Mult([-1,a],[d]), Mult([-1,b],[e])], Mult([d,e],[g])]

Proof.
  1 Mult([a,b],[c])                         c=(a*b)
  2 Mult([-1,a],[d])                        d=(-1*a)
  3 Mult([-1,b],[e])                        e=(-1*b)
  4 Mult([a,e],[f])     [T8,1,3]            f=(a*e)=(a*(-1*b))
  5 Mult([d,e],[g])     [T10,4,2]           g=(d*e)=((-1*a)*(-1*b))

Theorem T13.
[[Mult([a,b],[c]), Mult([-1,a],[d]), Mult([-1,b],[e]), Mult([d,e],[f])], Eq([f,c],[])]

Proof.
  1 Mult([a,b],[c])                         c=(a*b)
  2 Mult([-1,a],[d])                        d=(-1*a)
  3 Mult([-1,b],[e])                        e=(-1*b)
  4 Mult([d,e],[f])                         f=(d*e)=((-1*a)*(-1*b))
  5 Int([c],[])         [A2,1]              c:I
  6 Mult([-1,c],[g])    [A14,5]             g=(-1*c)=(-1*(a*b))
  7 Int([g],[])         [A2,6]              g:I
  8 Mult([-1,g],[h])    [A14,7]             h=(-1*g)=(-1*(-1*(a*b)))
  9 Eq([h,c],[])        [T7,6,8]            h=c
 10 Mult([a,e],[i])     [T8,1,3]            i=(a*e)=(a*(-1*b))
 11 Eq([i,g],[])        [T9,1,3,10,6]       i=g
 12 Mult([-1,i],[j])    [A19,2,4,10]        j=(-1*i)=(-1*(a*(-1*b)))
 13 Eq([j,f],[])        [A20,2,4,10,12]     j=f
 14 Eq([h,j],[])        [A4,12,11,8]        h=j
 15 Eq([h,f],[])        [A3,14,13]          h=f
 16 Eq([f,c],[])        [A3,9,15]           f=c

Theorem T14.
[[Lt([0,a],[]), Mult([-1,a],[b])], Lt([b,0],[])]

Proof.
  1 Lt([0,a],[])                            0<a
  2 Mult([-1,a],[b])                        b=(-1*a)
  3 Add([a,b],[c])      [A15,2]             c=(a+b)=(a+(-1*a))
  4 Eq([c,0],[])        [A16,2,3]           c=0
  5 Int([b],[])         [A2,2]              b:I
  6 Add([b,0],[d])      [A12,5]             d=(b+0)=((-1*a)+0)
  7 Eq([d,b],[])        [A13,6]             d=b
  8 Add([0,b],[e])      [A8,6]              e=(0+b)=(0+(-1*a))
  9 Eq([e,d],[])        [A9,6,8]            e=d
 10 Eq([e,b],[])        [A3,9,7]            e=b
 11 Lt([e,c],[])        [A26,1,8,3]         e<c
 12 Lt([e,0],[])        [A3,11,4]           e<0
 13 Lt([b,0],[])        [A3,12,10]          b<0

Theorem T15.
[[Lt([a,0],[]), Mult([-1,a],[b])], Lt([0,b],[])]

Proof.
  1 Lt([a,0],[])                            a<0
  2 Mult([-1,a],[b])                        b=(-1*a)
  3 Add([a,b],[c])      [A15,2]             c=(a+b)=(a+(-1*a))
  4 Eq([c,0],[])        [A16,2,3]           c=0
  5 Int([b],[])         [A2,2]              b:I
  6 Add([b,0],[d])      [A12,5]             d=(b+0)=((-1*a)+0)
  7 Eq([d,b],[])        [A13,6]             d=b
  8 Add([0,b],[e])      [A8,6]              e=(0+b)=(0+(-1*a))
  9 Eq([e,d],[])        [A9,6,8]            e=d
 10 Eq([e,b],[])        [A3,9,7]            e=b
 11 Lt([c,e],[])        [A26,1,3,8]         c<e
 12 Lt([0,e],[])        [A3,11,4]           0<e
 13 Lt([0,b],[])        [A3,12,10]          0<b

Theorem T16.
[[Lt([0,a],[]), Mult([a,a],[b])], Lt([0,b],[])]

Proof.
  1 Lt([0,a],[])                            0<a
  2 Mult([a,a],[b])                         b=(a*a)
  3 Int([a],[])         [A1,1]              a:I
  4 Mult([0,a],[c])     [T5,3]              c=(0*a)
  5 Eq([c,0],[])        [T6,4]              c=0
  6 Lt([c,b],[])        [A27,1,1,4,2]       c<b
  7 Lt([0,b],[])        [A3,6,5]            0<b

Theorem T17.
[[Lt([a,0],[]), Mult([a,a],[b])], Lt([0,b],[])]

Proof.
  1 Lt([a,0],[])                            a<0
  2 Mult([a,a],[b])                         b=(a*a)
  3 Int([a],[])         [A1,1]              a:I
  4 Mult([0,a],[c])     [T5,3]              c=(0*a)
  5 Eq([c,0],[])        [T6,4]              c=0
  6 Lt([c,b],[])        [A28,1,1,2,4]       c<b
  7 Lt([0,b],[])        [A3,6,5]            0<b

\end{lstlisting}

\vspace{5mm}
\noindent
\textbf{Notes.}

\begin{itemize}

\item In the theory of fields and commutative rings the identity $a=c-b$ follows trivially from the identity $c=a+b$.   
Theorems $T1$ and $T2$ highlight the difficulties associated with arithmetic on $\mb{I}$.
In $T1$ the existence of $c-b$ is established from the premise that $c=a+b$ exists.
$T2$ establishes the equality $a=c-b$.

\item Theorems $T5$ and $T6$ provide another example that highlights the difficulties associated with arithmetic on $\mb{I}$ where existence is not immediate.
In the theory of fields and commutative rings the existence of $0*a$ follows immediately from the closure of multiplication.
The proof of theorem $T5$ is rather a lengthy derivation dedicated just to the establishment that $0*a$ exists. 
The proof of the equality part $T6$ is almost identical to the existence part $T5$ but the statement $a:\mb{I}$ is redundent as a premise in $T6$. 

\item When combined, theorems $T16$ and $T17$ provide an example where disjunctions can be used.
We could have introduced the disjunction program $Lt([0,a],[~])~|~Lt([a,0],[~])$ and regarded the proofs of $T16$ and $T17$ as the parrallel operand programs that result from disjunction splitting.
The common conclusion $Lt([0,b],[~])$ of the operand programs can then be collapsed back onto the main proof program containing the disjunction.
Since we are avoiding the use of disjunctions in this paper the two cases $a<0$ and $a>0$ are treated separately.

\end{itemize}

\section{Concluding Remarks.}

There are several areas where future development of \emph{VPC} is envisaged.
Some of the areas that are mentioned below have already been investigated in some detail but have been omitted in this paper for purposes of brevity. 

\textbf{Analysis of program constructions as proofs.}
The construction rules that largely form the basis of the software package \emph{VPC} in its initial phase of development are presented as irreducible computable program extensions of higher order constructs.
The construction rules presented here can be regarded as axioms of the axiomatic system of program constructions as proofs.
However, they do not represent the full collection of axioms from which a comprehensive analysis of program constructions as proofs can be carried out.
At this stage it is understood that the foundation of the language relies heavily on the sublist rule and that much of the future work in this area needs to be focussed on extending the collection of auxiliary construction rules.

\textbf{Axiomatic systems.} Keeping with the initial motivation of this work, emphasis has been given to the study of machine arithmetic.
However, the formal language on which \emph{VPC} is based is not restricted to such applications and other axiomatic systems can be considered.
The process simply requires that the user define the specific axiomatic system of interest in the file, \emph{axiom.dat}, that is accessed by \emph{VPC} during proof construction.
There is a need to make a number of investigations using different axiomatic systems to assess the strengths and weaknesses of \emph{VPC} as a more general model of inference.

\textbf{Automated theorem proving.} Manual proof construction is a human/machine interactive process so that the user friendliness of \emph{VPC} needs to be enhanced.
Work is also in progress towards removing the human element for a fully automated theorem prover.
Automated theorem proving, at its most basic level, has severe drawbacks in that it suffers from combinatorial explosion.
Nevertheless, there is some hope in this area if one looks towards providing \emph{VPC} with some kind of strategy for goal oriented theorem searching.
This would involve the insertion of a module in \emph{VPC} whereby various strategies of automated decision making can be tried and tested.
One might expect that it is unlikely that a single multi-purpose strategy would be found that can be fixed into the \emph{VPC} software.
If this is the case then the human element remains a part of this process so that such a module would be a user specific strategy that \emph{VPC} would access before it is executed.
In other words the goal oriented automated theorem proving strategy would be part of the input data for \emph{VPC}. 
This suggests the initiation of a separate area of activity for the \emph{VPC} project that is one of examining various strategies for automated theorem proving.

\textbf{Disjunctions.} As has already been mentioned in a previous section, investigations into the most efficient way to deal with disjunctions is a work in progress.
The several approaches that have already been tried tend to lend themselves with differing weights towards manual and automated proof mining.
There is no inherint difficulty with disjunctions other than concerns of efficient storage and access of the simultaneous programs that arise from disjunction splitting and contraction.
This is also allied with extensions of the construction rules to include higher order axioms associated with noncomputable programs.  
These features are well advanced at this stage and are expected to be incoporated into the next version of \emph{VPC}. 

\textbf{Source code.} These preliminary notes attempt to describe the software package \emph{VPC} in a way that is independent of the higher order language upon which the source code is written.
The current version of \emph{VPC} is written in Fortran, but this reflects the author's familiarity with Fortran based on background experience rather than anything else.
There appears to be no reason why future versions of \emph{VPC} could not be written in other higher order languages.
These preliminary notes outline the features of the software package \emph{VPC} in its early phase of development and are not meant to serve as a users guide. 
A users guide for \emph{VPC} will be released when the abovementioned issues have been adequately addressed.

\vspace{5mm}

\appendix{\textbf{APPENDIX.}}

\begin{flushleft}

\section{Atomic integer programs.}

\vspace{5mm}
\textbf{Check type integer.}

\textbf{\textit{Syntax.}} $Int([a],[~])$.

\textbf{\textit{Program Type.}} $\mb{P}_{type}$.

\textbf{\textit{Type checks.}} $a:\mb{I}$.

\textbf{\textit{Description.}} $Int$ checks that the value assigned to $a$ has type $\mb{I}$.
$Int$ halts with an execution error if there is a type violation.

\vspace{5mm}
\textbf{Less than.}

\textbf{\textit{Syntax.}} $Lt([a,b],[~])$.

\textbf{\textit{Program Type.}} $\mb{P}_{type}$.

\textbf{\textit{Type checks.}} $a:\mb{I}$, $b:\mb{I}$, $a<b$.

\textbf{\textit{Calling programs.}} $Int([a],[~])$, $Int([b],[~])$. 

\textbf{\textit{Description.}} 
$Lt$ first checks that the values assigned to $a$ and $b$ are type $\mb{I}$.
It then checks that $a<b$.
$Lt$ halts with an execution error if there is a type violation.
Type violation includes the case where the assigned value of $a$ fails to be less than the assigned value of $b$.

\vspace{5mm}
\textbf{Numerical equality.}

\textbf{\textit{Syntax.}} $Eq([a,b],[~])$.

\textbf{\textit{Program Type.}} $\mb{P}_{type}$.

\textbf{\textit{Type checks.}} $a:\mb{I}$, $b:\mb{I}$, $a=b$.

\textbf{\textit{Calling programs.}} $Int([a],[~])$, $Int([b],[~])$.

\textbf{\textit{Description.}} 
$Eq$ first checks that the values assigned to $a$ and $b$ are type $\mb{I}$.
It then checks that $a=b$.
Here equality is in the sense of assigned values.
$Eq$ halts with an execution error if there is a type violation.
Type violation includes the case where the value assigned to $a$ fails to be equal to the value assigned to $b$.

\vspace{5mm}
\textbf{Not equal.}

\textbf{\textit{Syntax.}} $Neq([a,b],[~])$.

\textbf{\textit{Program Type.}} $\mb{P}_{type}$.

\textbf{\textit{Type checks.}} $a:\mb{I}$, $b:\mb{I}$, $a \neq b$.

\textbf{\textit{Calling programs.}} $Int([a],[~])$, $Int([b],[~])$.

\textbf{\textit{Description.}} 
$Neq$ first checks that the values assigned to $a$ and $b$ are type $\mb{I}$.
It then checks that $a \neq b$.
$Neq$ halts with an execution error if there is a type violation.
Type violation includes the case where the value assigned to $a$ is equal to the value assigned to $b$.

\vspace{5mm}
\textbf{Identity assignment.}

\textbf{\textit{Syntax.}} $Aid([a],[b])$.

\textbf{\textit{Program Type.}} $\mb{P}_{assign}$.

\textbf{\textit{Type checks.}} $a:\mb{I}$, $b:\mb{I}$.

\textbf{\textit{Calling programs.}} $Int([a],[~])$, $Int([b],[~])$.

\textbf{\textit{Assignment function.}} $b:=a$.

\textbf{\textit{Description.}} $Aid$ first checks that the value assigned to $a$ is type $\mb{I}$.
It then assigns to $b$ the value assigned to $a$, i.e. $b:=a$.
If $a:\mb{I}$ then the type check $b:\mb{I}$ is never violated.
$Aid$ returns the value $b$ as output provided that there are no type violations.
Otherwise it halts with an execution error.

\vspace{5mm}
\textbf{Addition.}

\textbf{\textit{Syntax.}} $Add([a,b],[c])$.

\textbf{\textit{Program Type.}} $\mb{P}_{assign}$.

\textbf{\textit{Type checks.}} $a:\mb{I}$, $b:\mb{I}$, $c:\mb{I}$.

\textbf{\textit{Calling programs.}} $Int([a],[~])$, $Int([b],[~])$, $Int([c],[~])$.

\textbf{\textit{Assignment function.}} $c:=a+b$.

\textbf{\textit{Description.}}
$Add$ first checks that the values assigned to $a$ and $b$ are type $\mb{I}$.
It then attempts to assign to $c$ the sum of $a$ and $b$, i.e. $c:=a+b$.
It then checks that $c:\mb{I}$.
This may fail if the sum $a+b$ is not contained within $\mb{I}$.
$Add$ returns the value $c$ as output provided that there are no type violations.
Otherwise it halts with an execution error.

\vspace{5mm}
\textbf{Multiplication.}

\textbf{\textit{Syntax.}} $Mult([a,b],[c])$.

\textbf{\textit{Program Type.}} $\mb{P}_{assign}$.

\textbf{\textit{Type checks.}} $a:\mb{I}$, $b:\mb{I}$, $c:\mb{I}$.

\textbf{\textit{Calling programs.}} $Int([a],[~])$, $Int([b],[~])$, $Int([c],[~])$.

\textbf{\textit{Assignment function.}} $c:=a*b$.

\textbf{\textit{Description.}}
$Mult$ first checks that the values assigned to $a$ and $b$ are type $\mb{I}$.
It then attempts to assign to $c$ the product of $a$ and $b$, i.e. $c:=a*b$.
It then checks that $c:\mb{I}$.
This may fail if the product $a*b$ is not contained within $\mb{I}$.
$Add$ returns the value $c$ as output provided that there are no type violations.
Otherwise it halts with an execution error.

\vspace{5mm}
\textbf{Division.}

\textbf{\textit{Syntax.}} $Div([a,b],[c])$.

\textbf{\textit{Program Type.}} $\mb{P}_{assign}$.

\textbf{\textit{Type checks.}} $a:\mb{I}$, $b:\mb{I}$, $c:\mb{I}$.

\textbf{\textit{Calling programs.}} $Int([a],[~])$, $Int([b],[~])$, $Int([c],[~])$.

\textbf{\textit{Assignment function.}} $c:=a/b$.

\textbf{\textit{Description.}}
$Div$ first checks that the values assigned to $a$ and $b$ are type $\mb{I}$.
It then attempts to assign to $c$ the value of $a$ divided by $b$, i.e. $c:=a/b$.
It then checks that $c:\mb{I}$.
This may fail if $b=0$ or if $b$ is not an integer multiple of $a$.
$Div$ returns the value $c$ as output provided that there are no type violations.
Otherwise it halts with an execution error.

\section{Atomic higher order programs.}

\vspace{5mm}
\textbf{Check type program.}

\textbf{\textit{Syntax.}} $Prog([p],[~])$.

\textbf{\textit{Program Type.}} $\mb{P}_{type}$.

\textbf{\textit{Type checks.}} $p:\mb{P}$.

\textbf{\textit{Description.}} $Prog$ checks that $p$ has the structure of a program, i.e.
$p$ has been assigned the value of a program list with representation of the form $P(x,y)=[ P_i(x_i,y_i) ]_{i=1}^n$, for some $n:\mb{I}_0$, satisfying all of the properties given in the definition of programs in Section 5.
$Prog$ halts with an execution error if there is a type violation.
The list length $n$ is not an input parameter since $Prog$ identifies $p$ as a list as well as determining its length.
The list length $n$ is an internally defined parameter of $Prog$ and is not returned as output.

\vspace{5mm}
\textbf{Check program sequential eqivalence.}

\textbf{\textit{Syntax.}} $Eqseq([p,q],[~])$.

\textbf{\textit{Program Type.}} $\mb{P}_{type}$.

\textbf{\textit{Type checks.}} $p:\mb{P}$, $q:\mb{P}$, $p \equiv q$.

\textbf{\textit{Calling programs.}} $Prog([p],[~])$, $Prog([q],[~])$.

\textbf{\textit{Description.}} $Eqseq$ first checks that $p:\mb{P}$ and $q:\mb{P}$.
It then checks that $p$ and $q$ are sequential equivalent, i.e. $p \equiv q$.
$Eqseq$ halts with an execution error if there is a type violation.
Type violation includes the case that $p$ and $q$ fail to be sequential equivalent.

\vspace{5mm}
\textbf{Check I/O equivalence.}

\textbf{\textit{Syntax.}} $Eqio([p,q],[~])$.

\textbf{\textit{Program Type.}} $\mb{P}_{type}$.

\textbf{\textit{Type checks.}} $p:\mb{P}$, $q:\mb{P}$, $p \thicksim q$.

\textbf{\textit{Calling programs.}} $Prog([p],[~])$, $Prog([q],[~])$.

\textbf{\textit{Description.}} $Eqio$ first checks that $p:\mb{P}$ and $q:\mb{P}$.
It then checks that $p$ and $q$ are I/O equivalent, i.e. $p \thicksim q$.
$Eqio$ halts with an execution error if there is a type violation.
Type violation includes the case that $p$ and $q$ fail to be I/O equivalent. 
 
\vspace{5mm}
\textbf{Check program sublist.}

\textbf{\textit{Syntax.}} $Sub([q,p],[~])$.

\textbf{\textit{Program Type.}} $\mb{P}_{type}$.

\textbf{\textit{Type checks.}} $p:\mb{P}$, $q:\mb{P}$, $q \subseteqq p$.

\textbf{\textit{Calling programs.}} $Prog([p],[~])$, $Prog([q],[~])$.

\textbf{\textit{Description.}} $Sub$ first checks that $p:\mb{P}$ and $q:\mb{P}$ and then checks that $q$ is a sublist of $p$, i.e. $q \subseteqq p$.
$Sub$ halts with an execution error if there is a type violation.
Type violation includes the case that $q$ is not a sublist of $p$.

\vspace{5mm}
\textbf{Check type computable program extension.}

\textbf{\textit{Syntax.}} $Cpe([p,c,s],[~])$.

\textbf{\textit{Program Type.}} $\mb{P}_{type}$.

\textbf{\textit{Type checks.}} $p:\mb{P}$, $c:\mb{P}$, $s:\mb{P}_{cpe}(p,c)$.

\textbf{\textit{Calling programs.}} $Prog([p],[~])$, $Prog([c],[~])$, $Prog([s],[~])$.

\textbf{\textit{Description.}} $Cpe$ first checks that $p:\mb{P}$, $c:\mb{P}$, $s:\mb{P}_{cpe}(p,c)$.
By the type association $s:\mb{P}_{cpe}$, it is understood that $s$ is gauranteed to be computable if $p$ is computable.
The hierachy of subtypes is $\mb{P}_{cpe}(p,c)<:\mb{P}$.
Axioms and theorems stored in the file \emph{axiom.dat} are automatically assigned the type $\mb{P}_{cpe}$.
Otherwise a program acquires the type $\mb{P}_{cpe}$ through an assignment by way of inference via the construction rules.
$Cpe$ halts with an execution error if there is a type violation, i.e. any one of $p$, $c$ and $s$ is not of type $\mb{P}$ and $s$ is not a program concatenation of $p$ and $c$ such that $s:\mb{P}_{cpe}(p,c)$.

\vspace{5mm}
\textbf{Computable program extension type assignment.}

\textbf{\textit{Syntax.}} $Acpe([p,c,s],[~])$.

\textbf{\textit{Program Type.}} $\mb{P}_{tassign}$.

\textbf{\textit{Type checks.}} $p:\mb{P}$, $c:\mb{P}$, $s:\mb{P}$, $s=[p,c]$.

\textbf{\textit{Calling programs.}} $Prog([p],[~])$, $Prog([c],[~])$, $Prog([s],[~])$.

\textbf{\textit{Description.}} $Acpe$ first checks that $p:\mb{P}$, $c:\mb{P}$, $s:\mb{P}$ and $s=[p,c]$.
If there are no type violations $Acpe$ then makes the assignment of subtype $s::\mb{P}_{cpe}(p,c)$.
The property $s:\mb{P}_{cpe}(p,c)$ states that $s$ is computable if $p$ is computable.
Once this subtype assignment is effected it is stored in memory so that $s$ retains the property $s:\mb{P}_{cpe}(p,c)$ whenever it is encountered in any following subprogam of the current program lists. 
$Acpe$ halts with an execution error if there is a type violation, i.e. any one of $p$, $c$ and $s$ is not of type $\mb{P}$ and $s$ is not a program concatenation of $p$ and $c$.

\vspace{5mm}
\textbf{Program list concatenation.}

\textbf{\textit{Syntax.}} $Conc([p,q],[r])$.

\textbf{\textit{Program Type.}} $\mb{P}_{assign}$.

\textbf{\textit{Type checks.}} $p:\mb{P}$, $q:\mb{P}$, $r:\mb{P}$.

\textbf{\textit{Calling programs.}} $Prog([p],[~])$, $Prog([c],[~])$, $Prog([r],[~])$.

\textbf{\textit{Assignment function.}} $r:=[p,q]$.

\textbf{\textit{Description.}} $Conc$ first checks that $p:\mb{P}$ and $q:\mb{P}$.
If succesful $Conc$ then assigns to $r$ the program concatenation of $p$ and $q$, i.e. $r:=[p,q]$.
$Conc$ then checks that $r:\mb{P}$.
This may fail if for instance the conditions $y_p \bigcap y_q =[~]$ and $x_p \bigcap y_q =[~]$ are not satisfied.
$Conc$ halts with an execution error if there is a type violation.

\end{flushleft}

\end{document}